\documentclass[12pt,a4paper]{article}
\usepackage[sumlimits]{amsmath}
\usepackage{amsfonts,amssymb,latexsym,eufrak,mathrsfs,graphicx,enumerate,bbm,bbold,epsfig,appendix}
\newcommand{\be}{\begin{equation}}
\newcommand{\ee}{\end{equation}}
\newcommand{\beq}{\begin{equation}}
\newcommand{\eeq}{\end{equation}}
\newcommand{\ba}{\begin{eqnarray}}
\newcommand{\ea}{\end{eqnarray}}
\newcommand{\bea}{\begin{eqnarray}}
\newcommand{\eea}{\end{eqnarray}}

\begin{document}
\baselineskip=15.5pt \pagestyle{plain} \setcounter{page}{1}


\def\del{{\partial}}
\def\vev#1{\left\langle #1 \right\rangle}
\def\cn{{\cal N}}
\def\co{{\cal O}}
\def\IC{{\mathbb C}}
\def\IR{{\mathbb R}}
\def\IZ{{\mathbb Z}}
\def\RP{{\bf RP}}
\def\CP{{\bf CP}}
\def\Poincare{{Poincar\`e}}
\def\tr{{\rm tr}}
\def\tp{{\tilde \Phi}}
\def\TL{\hfil$\displaystyle{##}$}
\def\TR{$\displaystyle{{}##}$\hfil}
\def\TC{\hfil$\displaystyle{##}$\hfil}
\def\TT{\hbox{##}}
\def\HLINE{\noalign{\vskip1\jot}\hline\noalign{\vskip1\jot}} 
\def\seqalign#1#2{\vcenter{\openup1\jot
  \halign{\strut #1\cr #2 \cr}}}
\def\lbldef#1#2{\expandafter\gdef\csname #1\endcsname {#2}}
\def\eqn#1#2{\lbldef{#1}{(\ref{#1})}%
\begin{equation} #2 \label{#1} \end{equation}}
\def\eqalign#1{\vcenter{\openup1\jot
    \halign{\strut\span\TL & \span\TR\cr #1 \cr
   }}}
\def\eno#1{(\ref{#1})}
\def\href#1#2{#2}
\def\half{{1 \over 2}}

\def\ads{{\it AdS}}
\def\adsp{{\it AdS}$_{p+2}$}
\def\cft{{\it CFT}}

\newcommand{\ber}{\begin{eqnarray}}
\newcommand{\eer}{\end{eqnarray}}
\newcommand{\beqar}{\begin{eqnarray}}
\newcommand{\cN}{{\cal N}}
\newcommand{\cO}{{\cal O}}
\newcommand{\cA}{{\cal A}}
\newcommand{\cT}{{\cal T}}
\newcommand{\cF}{{\cal F}}
\newcommand{\cC}{{\cal C}}
\newcommand{\cR}{{\cal R}}
\newcommand{\cW}{{\cal W}}
\newcommand{\eeqar}{\end{eqnarray}}
\newcommand{\eps}{\epsilon}
\newcommand{\pa}{\paragraph}
\newcommand{\pt}{\partial}
\newcommand{\de}{\delta}
\newcommand{\De}{\Delta}
\newcommand{\lb}{\label}


\newcommand{\oh}{\displaystyle{\frac{1}{2}}}
\newcommand{\dsl}
  {\kern.06em\hbox{\raise.15ex\hbox{$/$}\kern-.56em\hbox{$\partial$}}}
\newcommand{\id}{i\!\!\not\!\partial}
\newcommand{\as}{\not\!\! A}
\newcommand{\ps}{\not\! p}
\newcommand{\ks}{\not\! k}
\newcommand{\D}{{\cal{D}}}
\newcommand{\dv}{d^2x}
\newcommand{\Z}{{\cal Z}}
\newcommand{\N}{{\cal N}}
\newcommand{\Dsl}{\not\!\! D}
\newcommand{\Bsl}{\not\!\! B}
\newcommand{\Psl}{\not\!\! P}
\newcommand{\ZZ}{{\rm \kern 0.275em Z \kern -0.92em Z}\;}


\newcommand{\lbl}[1]{\label{eq:#1}}
\newcommand{ \rf}[1]{(\ref{eq:#1})}
\newcommand{\setl}{\setlength\arraycolsep{2pt}}

\newcommand{\noi}{\noindent}
\newcommand{\ra}{\rightarrow}
\newcommand{\Ra}{\Rightarrow}

\newcommand{\cd}{\bar{D}}
\newcommand{\cc}{\bar{C}}
\newcommand{\cB}{{\cal B}}
\newcommand{\cD}{{\cal D}}
\newcommand{\cG}{{\cal G}}
\newcommand{\cH}{{\cal H}}
\newcommand{\cK}{{\cal K}}
\newcommand{\cL}{{\cal L}}
\newcommand{\cM}{{\cal M}}
\newcommand{\cP}{{\cal P}}
\newcommand{\cS}{{\cal S}}
\newcommand{\cU}{{\cal U}}

\newcommand{\CF}{C_{\rm F}}
\newcommand{\Imm}{\mbox{\rm Im}}
\newcommand{\Ree}{\mbox{\rm Re}}
\newcommand{\Tr}{\mbox{\rm Tr}}
\newcommand{\Sp}{\mbox{\rm Sp}}
\newcommand{\Li}{\mbox{\rm Li}}
\newcommand{\MeV}{\mbox{\rm MeV}}
\newcommand{\GeV}{\mbox{\rm GeV}}
\newcommand{\fm}{\mbox{\rm fm}}

\newcommand{\with}{\mbox{\rm with}}
\newcommand{\while}{\mbox{\rm while}}
\newcommand{\annd}{\mbox{\rm and}}
\newcommand{\foor}{\mbox{\rm for}}
\newcommand{\oll}{\mbox{\rm all}}
\newcommand{\att}{\mbox{\rm at}}
\newcommand{\are}{\mbox{\rm are}}
\newcommand{\hc}{\mbox{\rm h.c.}}
\newcommand{\too}{\mbox{\rm to}}

\newcommand{\alphaQ}{\alpha_{\mbox{\rm eff}}(Q^2)}
\newcommand{\alphak}{\alpha_{\mbox{\rm eff}}(k_{E}^2)}
\newcommand{\alpham}{\alpha(\mu^2)}
\newcommand{\alphaq}{\alpha(Q^2)}
\newcommand{\al}{\alpha}
\newcommand{\als}{\alpha_{\mbox{\rm {\scriptsize s}}}}
\newcommand{\gs}{g_{\mbox{\rm {\scriptsize s}}}}
\newcommand{\muhad}{\mu_{\mbox{\rm {\scriptsize had.}}}}
\newcommand{\GF}{G_{\mbox{\rm {\tiny F}}}}
\newcommand{\MHA}{\mbox{\rm {\tiny MHA}}}

\newcommand{\qs}{\not \! q}
\newcommand{\xis}{\not \! \xi}
\newcommand{\gL}{\frac{1-\gamma_{5}}{2}}
\newcommand{\gR}{\frac{1+\gamma_{5}}{2}}
\newcommand{\gmut}{\mbox{$\tilde{\gamma}_{\mu}$}}
\newcommand{\smunut}{\mbox{$\tilde{\sigma}_{\mu\nu}$}}

\newcommand{\eff}{\mbox{\rm eff}}
\newcommand{\exxp}{\mbox{\rm exp.}}
\newcommand{\UV}{\mbox{\rm {\small UV}}}
\newcommand{\EM}{\mbox{\rm {\small EM}}}
\newcommand{\QCD}{\mbox{\rm {\footnotesize QCD}}}
\newcommand{\Lac}{\Lambda_{\chi}}
\newcommand{\msb}{\overline{\mbox{\rm\footnotesize MS}}}
\newcommand{\Lmsb}{\Lambda_{\overline{\mbox{\rm\footnotesize MS}}}}
\newcommand{\E}{\mbox{\rm {\tiny E}}}

\newcommand{\g}{\mbox{\bf g}}
\newcommand{\h}{\mbox{\bf h}}

\newcommand{\bP}{{\bf P}}
\newcommand{\bX}{{\bf X}}
\newcommand{\bQ}{{\bf Q}}
\newcommand{\Psls}{\not \! \bP}
\newcommand{\Qsls}{\not \! \bQ}
\newcommand{\ksls}{\not \! k}
\newcommand{\psls}{\not \! p}
\newcommand{\qsls}{\not \! q}
\newcommand{\pslsm}{\not \! p_{-}}
\newcommand{\pslsp}{\not \! p_{+}}

\newcommand{\stern}{\langle\bar{\psi}\psi\rangle}
\newcommand{\gew}{\mbox{\bf g}_{\mbox{\rm\footnotesize \underline{ew}}}}

\input epsf





\begin{titlepage}

\leftline{OUTP-06-10-P} \leftline{\tt hep-ph/0612010}

\vskip -.8cm

\vskip 1.5 cm

\begin{center}

{\LARGE Four-point functions and kaon decays in a minimal AdS/QCD model} \vskip .3cm

\vskip 1.cm

{\large Thomas Hambye$^{a,b,*}$, 
{Babiker Hassanain$^{a,\dag}$,
John March-Russell$^{a,\ddag}$ and \\
Martin Schvellinger$^{a,\P}$}} 
\vskip 0.6cm

{\it $^a$The Rudolf Peierls Centre for Theoretical Physics, \\
Department of Physics, University of Oxford. \\ 1 Keble Road,
Oxford, OX1 3NP, UK.} \\
$^\dag$ babiker@thphys.ox.ac.uk \\
$^\ddag$ jmr@thphys.ox.ac.uk    \\
$^\P$    martin@thphys.ox.ac.uk \\
\vskip 0.5cm

{\it $^b$IFT-UAM/CSIC, Facultad de Ciencias, \\ Universidad
Aut\'onoma de Madrid, \\ Cantoblanco, 28049 Madrid, Spain.} \\
$^*$ thomas.hambye@uam.es
\vspace{1.7cm}

\begin{abstract}

We study the predictions of holographic QCD for various observable
four-point quark flavour current-current correlators. The dual 5-dimensional 
bulk theory we consider is a
$SU(3)_L \times SU(3)_R$ Yang-Mills theory in a slice of $AdS_5$ spacetime
with boundaries.  Particular UV and IR boundary conditions encode the
spontaneous breaking of the dual 4D global chiral symmetry down to the $SU(3)_V$
subgroup.  We explain in detail how to calculate 
the 4D four-point quark flavour
current-current correlators using the 5D holographic theory, 
including interactions.
We use these results to investigate predictions of 
holographic QCD for the $\Delta I = 1/2$
rule for kaon decays and the $B_K$ parameter.  The results 
agree well in comparison
with experimental data, with an accuracy of $25 \%$ or better.
The holographic theory automatically includes the
contributions of the meson resonances to the four-point correlators.
The correlators agree well in the low-momentum and 
high-momentum limit, in comparison 
with chiral perturbation theory and perturbative QCD results, respectively.

\end{abstract}

\end{center}

\noindent

\end{titlepage}

\newpage

\tableofcontents

\newpage



\vfill

\section{Introduction}

The success of the perturbative description of QCD allows us to understand
the high energy behaviour of strong interactions above 1.5 GeV. On the other 
hand, chiral perturbation theory ($\chi$PT) describes well the physics of strong
interactions at low energy. In the intermediate region between both
regimes, the situation is much less clear since neither of these theories 
behave perturbatively. One interesting and potentially powerful new idea 
to gain access to the non-perturbative regime of QCD is holographic QCD, which
is based on the gauge/gravity duality \cite{Maldacena:1997re,Gubser:1998bc,
Witten:1998qj,Aharony:1999ti}. 

There are two kinds of holographic QCD dual models: 
there are 10D models based on string theory and supergravity 
\cite{Kruczenski:2003be}~-~\cite{Evans:2006ea}, including studies of deep inelastic scattering 
\cite{Polchinski:2001tt}~-~\cite{Brower:2006ea}, and
in addition, there are phenomenologically inspired 5D holographic dual models
\cite{Erlich:2005qh}~-~\cite{Shuryak:2006yx}.  In both approaches, the description
of confinement and chiral symmetry breaking 
has been tackled, and masses, decay constants, form factors and other properties 
of mesons have been calculated, yielding remarkably good agreement with experimental data. 
All of these estimates are based on two-point current correlators, which do not involve bulk
interactions and pertain to the low-lying mesons. Given these initial successes,
it is important that these holographic dual models of QCD are tested using processes that
go beyond the properties of two-point current correlators, and include 
interactions in the bulk of the 5D theory.
One such test is the computation in the 5D holographic theory of connected 4D four-point
flavour current correlators, which can be compared with experiment and, in certain limits,
with the results of chiral perturbation theory and perturbative QCD calculations.

In this paper, we shall focus exclusively on such four-point
flavour current correlators.  These correlators are crucial to the resolution of a 
long-standing problem in QCD: the $\Delta I=1/2$ rule, which we will describe 
in detail in later sections and briefly here.  In short, if one neglects CP-violating
effects, there  are two independent $K^0$ decays: $K^0 \rightarrow \pi^+ \pi^-$ and 
$K^0 \rightarrow \pi^0 \pi^0$. These two decays are combinations of 
$\Delta I =1/2$ and $\Delta I = 3/2$ isospin amplitudes, $A_0$ and $A_2$ 
respectively. Experimentally Re$A_0/$Re$A_2=22.2$, and the largeness of this
ratio is the $\Delta I = 1/2$ rule. In the chiral limit, these two amplitudes 
are generally expressed in terms of the $g_8$ and $g_{27}$ parameters (see 
for example \cite{Hambye:2003cy}), both of which depend on  
integrals over Euclidean momentum of certain four-current correlators. 
Our aim in this paper is to apply holographic QCD to calculate these
observables.\footnote{In a previous article \cite{we}, we briefly presented some of our
initial results on this subject. In the present work,
we significantly extend and improve upon our earlier study.}

The 4D theory we are trying to model is QCD with three massless quark flavours, 
possessing a global $SU(3)_L \times SU(3)_R$ symmetry which is spontaneously broken 
down to the vector subgroup via the quark condensate.\footnote{Note that for simplicity
throughout this paper we work in the chiral limit setting bare quark masses to zero, and 
ignore the anomalous and therefore explicitly broken $U(1)_{\rm axial}$ symmetry. 
We hope to return to these issues in a later publication.}  The AdS/CFT 
correspondence then immediately tells us that the dual 5D theory should be a 
Yang-Mills theory with $SU(3)_L \times SU(3)_R$ gauge group, 
with a bi-fundamental bulk scalar field to provide 
breaking of this symmetry. The gauge fields in 5D couple to the QCD flavour currents, 
whereas the bulk scalar couples to the bilinear quark operator. In previous models, the 
inclusion of a bulk scalar field allowed a comprehensive description of chiral 
symmetry breaking \cite{Erlich:2005qh,DaRold:2005zs,DaRold:2005vr,Katz:2005ir}.  
However, it is possible to take a limit of this theory where the entire description of
chiral symmetry breaking is encoded into the boundary 
conditions imposed on the gauge fields \cite{Hirn:2005nr,DaRold:2005zs} and the 
holographic theory contains only gauge fields in the bulk. This simplified holographic dual
model turns out to be a reasonable approximation \cite{Hirn:2005nr,DaRold:2005zs}, 
giving good results at least at the level of the two-point functions. The reason 
is that the condensate is an infrared (IR) effect, so that its influence 
can be modeled by an IR boundary condition. The complexity of the calculation of 
four-point current correlators in AdS/QCD means that this simpler 
form of holographic QCD, with only gauge fields in the bulk, 
provides an important starting point that can then be further refined.

Our results are encouraging for AdS/QCD.   As we discuss in 
detail in sections 5 and 6,
we find that at leading order in a low-momentum expansion, 
the behaviour of the relevant
correlators calculated in holographic QCD agrees with previous
calculations using chiral perturbation theory, while at high momentum we obtain
the behaviour predicted by perturbative QCD. In the intermediate 
region, the
momentum behaviour is governed by the exchange of meson resonances, and
a significant advantage of the holographic calculation is that it automatically and
consistently includes the contribution of the infinite tower of meson
resonances to the relevant correlators. Turning to a comparison with 
the experimental
data, the results of a fit of the holographic predictions agree well, 
with an accuracy of $25\%$ or better, which for quantities as difficult 
to calculate as the isospin amplitudes Re$A_0$ and Re$A_2$, is remarkable.   
Finally, we hope that the techniques developed here may be useful
for more general calculations of $n$-point global symmetry current correlators in many
AdS/CFT holographic dual models.

This paper is organised as follows:  In sections 2 and 3
we introduce the holographic QCD model that we use and present the relevant
5D propagators and interactions.   In section 4 we discuss
the $\Delta I=1/2$ observables from the viewpoint of QCD, as well as a related 
but simpler observable $\hat{B}_K$, parameterising $K^0-\bar{K}^0$ mixing. 
We shall also define the nature of the four-current correlators that we 
calculate using the holographic dual model, and the dependence of the 
parameters $g_8$, $g_{27}$ and $\hat{B}_K$ on these correlators.
We also review the $\chi$PT predictions for the various observables.
In section 5 we discuss the philosophy of the calculation and present the sum
of the 5D Witten diagrams relevant for four-point functions.  Section 6 contains
our numerical results, and our conclusions are given in section 7,
where we also briefly address the limitations of the model and possible avenues of improvement.
Finally, four appendices contain technical details of the holographic calculation.

Before we start upon our analysis we think it may be useful to offer the readers
a ``road map'' to follow the contents of this paper, depending on their particular
interests. For readers interested in the AdS/QCD model and calculations of $n$-point correlators, 
and in particular four-point current correlators, sections 2, 3, and 5, supplemented 
with appendices A, B, C and D are recommended. For those 
interested in the physics of the kaon decays from $\chi$PT
and perturbative QCD, section 4 is relevant. Readers interested in 
the comparison of the holographic calculation with the experimental results 
are directed towards section 6.

\section{The 5D Holographic Model}
\lbl{model}

Motivated by the AdS/CFT correspondence, and following on from the work of
Refs.\cite{Erlich:2005qh,DaRold:2005zs,Hirn:2005nr,DaRold:2005vr,Katz:2005ir},
we consider a 5D bulk theory
defined in a constant curvature spacetime with the minimal field content
as to describe current-current correlators in QCD.
The spacetime metric is that of AdS$_5$ space
\be
ds^2 = a^2(z) \, (\eta_{\mu\nu} \, dx^\mu dx^\nu - dz^2) \, ,
\ee
where $a(z)=L/z$, $L$ being the curvature scale of the anti-de-Sitter
space. The 5th-dimensional coordinate $z$ holographically represents
the energy scale of the 4D theory.  We take $z$ to extend from 
a UV boundary at $z=L_0$ to an IR boundary at $z=L_1>L_0$.

We are only interested in spin-1 4D operators such as 
${\bar q}_L \gamma^\mu t^a q_L$ and ${\bar q}_R \gamma^\mu t^a q_R$,
where $q$ can be $u$, $d$ and $s$ quarks.  Using the well-known
AdS/CFT relation between the dimension of such spin-1 boundary-theory operators 
$\Delta$ and the mass of the bulk vector fields $m_5$,  
$(\Delta + 1) (\Delta - 3)=m_5^2$, we find that $\Delta=3$ gives $m_5=0$. 
We consider the chiral limit of QCD where the quarks are massless, so that 
global flavour currents are conserved and the boundary symmetry group is 
a global $SU(3)_L \times SU(3)_R$. 

The rules of the holographic correspondence then tell us that
the bulk theory is a pure 5D Yang-Mills with {\it gauge} group $SU(3)_L \times SU(3)_R$.
The boundary conditions on the UV brane $z=L_0$ are such that the zero
modes of the gauge fields in the $\mu$ directions are eliminated, so that no
massless 4D gauge symmetry survives.
The Lagrangian is given by \cite{Erlich:2005qh,DaRold:2005zs,Hirn:2005nr}
\be \label{Lagrangian}
{\cal {L}}_{5D} = \sqrt{g} \, M_5 \, Tr \left( -\frac{1}{4} \, L_{MN} \,
L^{MN} - \frac{1}{4} \, R_{MN} \, R^{MN}
         \right) \, .
\ee 
The scale $M_5$ is some yet undetermined mass scale, $g$ is the determinant of
the metric and  $M=(\mu, 5)$, where $\mu=1, \cdot \cdot \cdot, 4$.  
The trace $Tr$ is taken over the gauge group indices.

We have $L_M = L_M^a \, T^a$ and similarly for $R_M$, where $T^a$ are the 
Hermitian generators for the Lie Algebra of the $SU(3)_L$ and $SU(3)_R$ 
groups, satisfying the following commutation relations and normalisations
\begin{eqnarray}
[T^a,T^b]=if^{abc}T^c & \textrm{and} & Tr[T^aT^b]=\delta^{ab} \, .
\end{eqnarray}
The $f^{abc}$'s are real and anti-symmetric in this basis. We
write the following expressions for the gauge field strengths
\begin{eqnarray}
L_{MN}&=&\partial_ML_N-\partial_NL_M-i[L_M,L_N] \, , \\
R_{MN}&=&\partial_MR_N-\partial_NR_M-i[R_M,R_N] \, ,
\end{eqnarray}
which give us the following relation 
\be
L_{MN}^a=\partial_ML_N^a-\partial_NL_M^a + f^{abc}L_M^bL_M^c \, ,
\ee
and similarly for $R_{MN}$.

We wish to work with the vector and axial-vector combinations of these gauge fields, 
so we define
\ba
V_M &=& \frac{1}{\sqrt{2}} \, (L_M + R_M) \, , \\
A_M &=& \frac{1}{\sqrt{2}} \, (L_M - R_M) \, .
\ea
The reason behind this choice is simple: the spontaneous breaking of chiral symmetry 
mixes the $L_M$ and $R_M$ gauge fields at the quadratic level. Therefore, the 
choice of basis as vector and axial-vector rather than left- and right-handed can 
be viewed as a diagonalisation of the equations of motion. Of course, at the cubic and 
quartic interaction level, there is mixing between $V_M$ and $A_M$, a fact which 
is integral to the calculation presented in this article. 

We can then express the Lagrangian above entirely in terms of vector and axial-vector fields.
To eliminate the mixing between $V_\mu$ and $V_5$ and between 
$A_\mu$ and $A_5$, we need to include the following $R_\xi$ gauge fixing terms
\ba
{\cal {L}}^V_{GF} &=&  - \frac{M_5 \, a}{2 \xi} \, Tr \left(\eta^{\mu\nu} \partial_\mu V_\nu -
\frac{\xi}{a} \, \partial_5(a V_5) \right)^2  \, , \\
{\cal {L}}^A_{GF} &=&  - \frac{M_5 \, a}{2 \xi} \, Tr \left(\eta^{\mu\nu} \partial_\mu A_\nu -
\frac{\xi}{a} \, \partial_5(a A_5) \right)^2 \, ,
\ea
where $\xi$ is the gauge parameter. 

We will go into unitary gauge in what follows, taking the limit $\xi \to \infty$ at 
the appropriate stages of the calculations. Note that this will have different effects on the 
vector and axial-vector sectors, due to the different boundary conditions we impose, as 
explained below. In the next section, we present the full Lagrangian in terms of $V_M$ and $A_M$,
and discuss the IR boundary conditions at $z=L_1$, which correspond to spontaneously
breaking the global $SU(3)_L \times SU(3)_R$ symmetry down to its $SU(3)_V$ subgroup.

\section{Propagators and Interactions}
\lbl{propagators}

As we later discuss in detail in section 4, we wish to calculate certain four-point
current correlators involving two 
left-handed and two right handed currents. We can expand the relevant desired
correlator in terms of the vector and axial-vector currents $J_V^\mu$ and $J_A^\mu$, 
so that we can use the bulk Lagrangian in terms of the vector and 
axial-vector fields. 

According to the AdS/CFT correspondence, the boundary values of $V_\mu$ 
and $A_\mu$ are classical sources coupling to $J_V^\mu$ and $J_A^\mu$,
respectively. In order to calculate tree-level $n$-point functions for 
the currents, we need to solve the bulk equations 
of motion for the vector and axial-vector fields, substitute back 
into the action and treat this as the generating functional of the 
boundary theory. One thus has
\begin{equation}
\left<e^{\int \mathrm{d}^4x J_V^\mu(x) v_\mu(x) + J_A^\mu(x) a_\mu(x) } \right> 
= e^{-S_{AdS}} \, ,
\end{equation}
where, $S_{AdS}$ is the Euclidean classical bulk action calculated with 
$V_\mu\vert_{UV}=v_\mu$ and $A_\mu\vert_{UV}=a_\mu$. We therefore need the 
bulk-to-bulk and bulk-to-boundary propagator for each of the gauge fields in the AdS 
field theory \cite{Witten:1998qj,Freedman}. The former allows us to construct the 
solution to the equations of motion from the interactions in the bulk 
of the AdS space, and the latter allows the construction of the solution 
of the equations of motion from the UV boundary value of the field. 
Green's second theorem gives us a straightforward relation between 
the two types of propagators.

The procedure of finding the on-shell AdS action subject to 
certain boundary values of the fields can equivalently be formulated 
in terms of Witten diagrams, where one uses the vertices of the bulk 
theory to construct all the allowed Feynman diagrams connecting 
the boundary operators. The ingredients are the propagators 
and the vertices, as calculated from the bulk Lagrangian. In this 
section, we describe how to calculate the propagators.  The 
vertices are simply derived from the full Lagrangian given in
Eqs.(\ref{boundaryL})-(\ref{fullL}). We first justify 
our choices of boundary conditions for the various fields.

Using the variational principle, the bulk equations of motion can 
be derived, along with the constraints that must be obeyed 
by any set of consistent boundary conditions. 
The UV boundary conditions on the bulk-to-bulk propagators 
can be chosen to be null Dirichlet for both the vector and axial-vector sectors. 
The IR boundary conditions distinguish the sectors, and allow 
chiral symmetry breaking ($\chi$SB) to be implemented into our model, by imposing null 
Neumann and null Dirichlet conditions on the vector and axial-vector 
sectors, respectively. This choice can be understood via an 
elegant argument: one can consider a bi-fundamental scalar 
living on the IR brane which acquires a vacuum expectation value. 
The effect of this on the boundary conditions is simple: it does not 
affect the vector sector, but changes the boundary conditions on the 
axial-vector fields from Neumann to mixed. This breaks the chiral 
symmetry spontaneously, and in theory we have one parameter to play 
with, analogous to the size of the quark condensate. Now, imagine 
removing the brane scalar from the theory by allowing its mass to 
go to infinity. The boundary condition on the axial-vector fields 
is now found to be a null Dirichlet condition. We also lose the parameter 
that allows us to tune the size of the symmetry breaking relative to the scale
$1/L_1$, which is set by the IR brane position.

The final requirement is of course to account for the pions, 
which form a massless pseudo-scalar octet. We do this as follows: we 
impose warped Neumann boundary conditions on the $A_5$ field on 
both branes. This guarantees that even after going to unitary gauge in 
the axial sector, a zero mode remains in $A_5$ and cannot be gauged
away. In the vector sector, we impose Dirichlet conditions on both 
branes for $V_5$, so that going into unitary gauge here removes 
$V_5$ from the theory.

\subsection{Interaction terms}
\label{totalL}

Here, we display the full boundary Lagrangian and 
bulk Lagrangian. Note that in the quadratic part, we have taken the 
limit $\xi \to \infty$ inside the differential operator acting on $A_\mu$ and 
$V_\mu$, but not in the $A_5$ operator. The quadratic part of the Lagrangian 
is used to calculate the propagators as shown below. The interaction 
Lagrangian provides the vertices needed to construct the Witten diagrams 
relevant for the calculation of any given $n$-point boundary current 
correlator. The full 5D Lagrangian can be written as a sum of the following 
terms, where the trace over the gauge indices is implicit 
\begin{equation}
\mathcal{L}_{boundary}=\left[\frac{M_5L}{2z}\eta^{\mu\nu}
\left(V_\nu\partial_5 V_\mu+A_\nu\partial_5 A_\mu
-2A_\mu\partial_\nu A_5\right)\right]_{L_0}^{L_1} ,
\label{boundaryL}
\end{equation}
\begin{eqnarray}
\mathcal{L}_{quadratic}&=&\frac{M_5L}{2z}
\left[V_\mu\left(\partial^2\eta^{\mu\nu}-z\partial_z\left(\frac{1}{z}\partial_z\right)
\eta^{\mu\nu}-\partial^{\mu}\partial^{\nu}\right)V_\nu\right. \nonumber \\
& & A_\mu\left(\partial^2\eta^{\mu\nu}-z\partial_z\left(\frac{1}{z}\partial_z\right)
\eta^{\mu\nu}-\partial^{\mu}\partial^{\nu}\right)A_\nu  \nonumber \\
& & \left. +A_5(-\partial^2)A_5
+\xi A_5\partial_z\left( z\partial\left(\frac{1}{z}A_5\right)\right)\right] ,
\end{eqnarray}
\begin{equation}
\mathcal{L}_{VAA_5,VA_5A_5}=-i\frac{M_5L}{\sqrt{2}z}\eta^{\mu\nu}
\left(\partial_5 V_\mu[A_5,A_\nu]+\partial_5 A_\mu[A_5,V_\nu]
+\partial_\mu A_5[V_\nu,A_5]\right) , 
\end{equation}
\begin{equation}
\mathcal{L}_{3-vector}=i\frac{M_5L}{\sqrt{2}z}\eta^{\mu\rho}\eta^{\nu\sigma}
\left(\partial_\mu V_\nu[A_\rho,A_\sigma]+\partial_\mu A_\nu[V_\rho,A_\sigma]
+\partial_\mu A_\nu[A_\rho,V_\sigma]+\partial_\mu V_\nu[V_\rho,V_\sigma]\right) ,
\end{equation}
\begin{eqnarray}
\mathcal{L}_{4-vector}&=&\frac{M_5L}{4z}\eta^{\mu\rho}\eta^{\nu\sigma}
(V_\mu V_\nu[V_\rho,V_\sigma]+A_\mu A_\nu[A_\rho,A_\sigma]
+4V_\mu V_\nu A_\rho A_\sigma-2V_\mu V_\nu A_\sigma A_\rho \nonumber\\
& & +2V_\mu A_\nu V_\rho A_\sigma-2V_\mu A_\nu A_\sigma V_\rho
-A_\mu V_\rho A_\nu V_\sigma-V_\mu A_\rho V_\nu A_\sigma) , 
\end{eqnarray}
\begin{equation}
\mathcal{L}_{VVA_5A_5,AAA_5A_5}=-\frac{M_5L}{2z}\eta^{\mu\nu}
\left(V_\mu A_5[V_\nu,A_5]+A_\mu A_5[A_\nu,A_5]\right) . \label{fullL}
\end{equation}
Examining the boundary Lagrangian in Eq.(\ref{boundaryL}), one immediately sees the familiar terms that 
are bilinear in $V_\mu$ and $A_\mu$. These are responsible for the emission 
of vector and axial-vector resonances by the boundary theory current. The 
unusual term here is the mixing term between $A_\mu$ and $A_5$, which says that 
a boundary axial current can emit an $A_5$ particle. This term survives 
the application of the boundary conditions, and is in fact of paramount 
importance in the satisfaction of the Ward identities. This can be seen from 
a calculation of the axial current two-point function as described in appendix C.
The Ward identity requires 
this correlator to be transverse, but this is only achieved by the AdS/CFT computation 
if one takes into account a diagram where the $A_5$ field is emitted by one 
current and absorbed by the other.

Finally, it is clear that the inclusion of the higher dimensional 
operators $Tr(L_{MN}^3+R_{MN}^3)$ and $Tr(L_{MN}^4+R_{MN}^4)$ in the Lagrangian of Eq.(\ref{Lagrangian}) 
results in other three-boson and four-boson vertices. However, one can show that they 
are sub-leading in the large $M_5L$ or, equivalently, the large $N_c$ limit, once one 
recognizes that $M_5L$ goes like $N_c$ parametrically, as in Refs.\cite{Erlich:2005qh,DaRold:2005zs}.
The argument goes as follows: the Lagrangian in five dimensions has the schematic form 
given by
\begin{equation}
\mathcal{L}_{5D}=M_5 L \left[\frac{\sqrt{g}}{L}Tr(F^2) + c_1 \frac{\sqrt{g}}{L}  \frac{Tr(F^3)}{M_5^2} 
+ c_2 \frac{\sqrt{g}}{L} \frac{Tr(F^4)}{M_5^4} + \cdots \right] \, ,
\end{equation}
where $Tr(F^2)$ represents the leading terms of the gauge field theory as written in 
Eq.(\ref{Lagrangian}), i.e. $Tr(L_{MN}L^{MN}+R_{MN}R^{MN})$, and the other terms signify 
higher dimensional operators, an example being $Tr(g^{MS}g^{NQ}g^{PR}L_{MN}L_{SP}L_{QR}+L \to R)$.
Now, when we write this in terms of the fields $A_\mu,V_{\mu}$ and $A_5,V_5$, we must remember 
that each factor of field strength $F$ comes in with a factor of $g^{MN}$. A factor of $g^{MN}$ brings 
with it a factor $z^2/L^2$. Recall also that $\sqrt{g}=(L/z)^5$. Thus, we can schematically write:
\begin{eqnarray}
L_{5D}&=&M_5 L \left[\frac{1}{z}Tr(F_{dd}F_{dd}) + c_1 \frac{z}{(M_5 L)^2} Tr(F_{dd}F_{dd}F_{dd})
          \right. \nonumber \\ 
      && \left.+ c_2 \frac{z^3}{(M_5 L)^4} Tr(F_{dd}F_{dd}F_{dd}F_{dd}) + \cdots \right] \, ,
\end{eqnarray}
where $F_{dd}$ simply means $L_{MN},R_{MN}$, i.e. the field strength with lower Lorentz indices. 
More generally, a gauge invariant operator with $n$ factors of $F$ will have a coefficient 
that goes like $c_{n-2}z^{2n-5}/(M_5 L)^{2(n-2)}$, where all the $c$ factors are of order one. 
This means that, in the large $N_c$ limit, the contribution from these operators is sub-leading to 
that from the term $Tr(L_{MN}L^{MN}+R_{MN}R^{MN})$, as claimed.

\subsection{$SU(3)_V$ sector propagators}

The vector {\it bulk-to-bulk} propagator has a transverse and longitudinal
part, because we are working in the gauge $V_5=0$ (unitary gauge).
The value of $V_\mu$ at the UV boundary is the classical source to
which the 4D current $J_V^\mu$ couples. On the IR boundary, we
impose a Neumann condition on the vector field. We write
(following \cite{Randall:2001gb})
\begin{equation}
\langle V^\mu V^\nu \rangle=
-iG^V_p(z,z')\left(\eta^{\mu\nu}-\frac{p^\mu p^\nu}{p^{2}}\right)
-iG^V_0(z,z')\left(\frac{p^\mu p^\nu}{p^{2}}\right) \, .
\end{equation}
Note that we are working in Lorentz indices, so we lower and raise
these indices with the 4D Poincar\'e metric $\eta^{\mu \nu}$. In Fourier 
space in the $x^\mu$ directions, but position space for the 5th direction,
the propagators solve the equation
\begin{equation}
\left(\partial_z^2-\frac{1}{z}\partial_z+p^2
\right)G^V_p(z,z')=\frac{z}{M_5L}\delta(z-z') \, .
\end{equation}
In addition, $G^V_0(z,z')$ solves the same equation with $p$ set
to zero. The boundary conditions on $G^V_p(z,z')$ are Dirichlet on
the UV brane and Neumann on the IR brane, so that
\begin{eqnarray}
G^V_p(z,z')\big\vert_{z=L_0}=0 \, ,\\
\partial_z G^V_p(z,z')\big\vert_{z=L_1}=0 \, .
\end{eqnarray}
From Green's second theorem, the {\it bulk-to-boundary} propagator is 
defined by the following limit
\begin{equation}
\langle V^\mu V^\nu \rangle\bigg\vert_{\partial
ADS}(z')=-\frac{M_5L}{z}\partial_z\langle V^\mu
V^\nu\rangle\Big\vert_{z=L_0} \, ,
\end{equation}
where
\begin{equation}
\langle V^\mu V^\nu \rangle\bigg\vert_{\partial ADS}(z')=
-iK^V_p(z')\left(\eta^{\mu\nu} -\frac{p^\mu
p^\nu}{p^{2}}\right)-iK^V_0(z')\left(\frac{p^\mu
p^\nu}{p^{2}}\right) \, .
\label{Vbulkboundary}
\end{equation}
The solutions are given by \cite{Erlich:2005qh,DaRold:2005zs,Hirn:2005nr}
\begin{eqnarray}
G^V_p(z,z')_{z<z'}=\frac{\pi
zz'}{2M_5L(AD-BC)}[A\mathcal{J}_1(pz)+B\mathcal{Y}_1(pz)]
[C\mathcal{J}_1(pz')+D\mathcal{Y}_1(pz')] \, , \\
G^V_p(z,z')_{z>z'}=\frac{\pi
zz'}{2M_5L(AD-BC)}[A\mathcal{J}_1(pz')+B\mathcal{Y}_1(pz')]
[C\mathcal{J}_1(pz)+D\mathcal{Y}_1(pz)] \, ,
\end{eqnarray}
where
\begin{eqnarray}
A=&-\mathcal{Y}_1(pL_0) , \quad &\textrm{} \quad
C=-\mathcal{Y}_0(pL_1) \, , \nonumber  \\
B=&\mathcal{J}_1(pL_0) , \quad &\textrm{} \quad
D=\mathcal{J}_0(pL_1)  \, ,
\end{eqnarray}
and
\begin{eqnarray}
G^V_0(z,z')_{z<z'}=-\frac{1}{2M_5L}(z^2-L_0^2)  \, ,  \\
G^V_0(z,z')_{z>z'}=-\frac{1}{2M_5L}(z'^2-L_0^2) \, .
\end{eqnarray}
Here $\mathcal{J}$ and $\mathcal{Y}$ are Bessel functions 
of the first and second kind in the conventions of Ref.\cite{AS}.
From these bulk-to-bulk propagators, we find that the
bulk-to-boundary propagators are given by
\begin{eqnarray}
K^V_p(z') & = &
-\frac{z'}{L_0}\frac{[C\mathcal{J}_1(pz')+D\mathcal{Y}_1(pz')]}{[AD-BC]}
\, ,\\
K^V_0(z') & = &  1 \, .
\end{eqnarray}
Note that in calculating the bulk-to-boundary propagator, we use
the bulk-to-bulk propagator for $z<z'$, so that
\begin{eqnarray}
K^V_p(z') & = & -\frac{M_5L}{z}\partial_z
G^V_p(z,z')_{z<z'}\Big\vert_{z=L_0}  \, , \\
K^V_0(z') & = & -\frac{M_5L}{z}\partial_z
G^V_0(z,z')_{z<z'}\Big\vert_{z=L_0}  \, .
\end{eqnarray}

\subsection{$SU(3)_A$  sector propagators}

We here list the propagators for $A_\mu$ and $A_5$. As explained above, the IR
boundary conditions in this sector are chosen so that the $SU(3)_L \times SU(3)_R$
global chiral symmetry is spontaneously broken:
\begin{eqnarray}
G^A_p(z,z')\big\vert_{z=L_0}=0 \, ,\\
G^A_p(z,z')\big\vert_{z=L_1}=0 \, .
\end{eqnarray}
Similarly to the $SU(3)_V$ sector we define
\begin{equation}
\langle A^\mu A^\nu \rangle=
-iG^A_p(z,z')\left(\eta^{\mu\nu}-\frac{p^\mu p^\nu}{p^{2}}\right)
-iG^A_0(z,z')\left(\frac{p^\mu p^\nu}{p^{2}}\right) \, .
\end{equation}
The equation to be solved is the same as in the $SU(3)_V$ sector. The results are
\cite{Erlich:2005qh,DaRold:2005zs,Hirn:2005nr}
\begin{eqnarray}
G^A_p(z,z')_{z<z'}=\frac{\pi
zz'}{2M_5L(A\cd -B\cc)}[A\mathcal{J}_1(pz)+B\mathcal{Y}_1(pz)]
[\cc \mathcal{J}_1(pz')+\cd \mathcal{Y}_1(pz')] \, , \\
G^A_p(z,z')_{z>z'}=\frac{\pi
zz'}{2M_5L(A\cd -B\cc)}[A\mathcal{J}_1(pz')+B\mathcal{Y}_1(pz')]
[\cc \mathcal{J}_1(pz)+\cd \mathcal{Y}_1(pz)] \, ,
\end{eqnarray}
where $A$ and $B$ are as for the $SU(3)_V$ sector, and 
\begin{equation}
\cc =-\mathcal{Y}_1(pL_1) , \quad \quad \cd =\mathcal{J}_1(pL_1) \, .
\end{equation}
This gives us
\begin{eqnarray}
G^A_0(z,z')_{z<z'}=-\frac{1}{2M_5L}(z^2-L_0^2) \, \frac{z'^2-L_1^2}
{L_0^2-L_1^2} \, , \\
G^A_0(z,z')_{z>z'}=-\frac{1}{2M_5L}(z'^2-L_0^2) \,
\frac{z^2-L_1^2} {L_0^2-L_1^2} \, .
\end{eqnarray}
From these propagators, we obtain the bulk-to-boundary propagators
as for the $SU(3)_V$ case, to find that
\begin{eqnarray}
K^A_p(z') & = &
-\frac{z'}{L_0}\frac{[\cc \mathcal{J}_1(pz')+\cd \mathcal{Y}_1(pz')]}{[A\cd -B\cc ]}\,
, \\ K^A_0(z') & = & \frac{z'^2-L_1^2}{L_0^2-L_1^2} \, .
\end{eqnarray}
We also note that the current-current correlator as $p\to 0$ is
now given by (see appendix C)
\begin{equation}
\Pi_A(p^2)\vert_{p=0}= F_{\pi}^2 =\frac{2M_5L}{L_1^2-L_0^2}.
\end{equation}
This is a direct consequence of the IR boundary condition. 
The $A_5$ propagator in this gauge is simply given by
\begin{equation}
\langle A_5 A_5 \rangle= -iG^5_p(z,z') ,
\end{equation}
where $G^5_p(z,z')$ is the limit as $\xi \to \infty$ of the solution to 
the equation
\begin{equation}
\left(\xi\partial_z^2-\xi\frac{1}{z}\partial_z+\xi\frac{1}{z^2}+p^2
\right)G^5_p(z,z')=-\frac{z}{M_5L}\delta(z-z') \, ,
\end{equation}
with the boundary conditions
\begin{eqnarray}
\partial_z(aG^5_p(z,z'))\big\vert_{z=L_0}=0 \, ,\\
\partial_z(aG^5_p(z,z'))\big\vert_{z=L_1}=0 \, ,
\end{eqnarray}
giving 
\footnote{Note the factor 2 difference between the $A_5$ propagator here and the 
(incorrect) one used in our previous paper \cite{we}.}
\begin{equation}
G^5_p(z,z')=\frac{-2zz'}{M_5L[L_1^2-L_0^2]}\left(\frac{1}{p^2}\right) .
\end{equation}

\section{Four-point Observables}
\lbl{int}

\subsection{The $\Delta I = 1/2$ rule}

Neglecting CP violation effects, there are two independent 
$K \rightarrow \pi \pi$ decay amplitudes, 
$K^0 \rightarrow \pi^+ \pi^-$ and $K^0 \rightarrow \pi^0 \pi^0$. 
These amplitudes can be written in terms of the $\Delta I=1/2$ amplitude 
$A_0$ and the $\Delta I = 3/2$ amplitude $A_2$ as
\begin{eqnarray}
A(K^0 \rightarrow \pi^+ \pi^-)&=& A_0 e^{i \delta^0} + 
\sqrt{1/2} A_2 e^{i \delta_2},
\\
A(K^0 \rightarrow \pi^0 \pi^0)&=& A_0 e^{i \delta^0} - 
\sqrt{2} A_2 e^{i \delta_2}.
\end{eqnarray}
The measured values of these amplitudes are
\begin{equation}
\mbox{Re}A_0=2.72 \cdot 10^{-4}~{\mbox{MeV}},  \quad  
\quad  {\mbox{Re}A_2}=1.22 
\cdot 10^{-5}~{\mbox{MeV}} \,,
\label{gdata}
\end{equation}
which gives
\begin{equation}
\frac{1}{\omega}\,\equiv\,
\frac{\mbox{Re}A_0}{\mbox{Re}A_2}\,\equiv\,
\frac{\mbox{Re}(K
\rightarrow(\pi\pi)_{I=0})}{\mbox{Re}(K\rightarrow
(\pi \pi)_{I=2})}\,=\,22.2\,.
\end{equation}
The large value of Re$A_0/$Re$A_2$ is the so called $\Delta I = 1/2$ rule. 

In the following, we use holographic QCD to calculate $\mbox{Re}A_0$ 
and $\mbox{Re}A_2$ in the chiral limit. In this limit, at order $p^2$ 
in the chiral counting, all the $\Delta S = 1$ transitions can be obtained
from the standard $\Delta S = 1$ effective Lagrangian, involving the usual 
$g_8$ and $g_{27}$ coupling constants (neglecting the 
small electromagnetic contribution, see 
for example ~\cite{KMW,Hambye:2003cy})
\be\label{ewcl}
\cL_{\eff}^{\Delta S=1}=-\frac{\GF}{\sqrt{2}}V_{ud}V_{us}^*\left[
g_{8}\cL_{8}+g_{27}\cL_{27} \right]\,,
\ee
where
\begin{equation}
\cL_8=\sum_{i=1,2,3}(\cL_{\mu})_{2i}\ (\cL^{\mu})_{i3}\quad\annd\quad
\cL_{27}=\frac{2}{3}(\cL_{\mu})_{21}\ (\cL^{\mu})_{13}+
(\cL_{\mu})_{23}\ (\cL^{\mu})_{11}\,,
\end{equation}
with
\begin{equation}
\cL_{\mu}=-iF_{\pi}^2\ U(x)^{\dagger}D_{\mu} U(x)\,,
\end{equation}
and $V_{ud}=0.974$, $V_{us}=0.224$. The pion decay constant $F_\pi$ 
is taken in the chiral limit, where the masses of the $u$, $d$ and $s$ quarks 
are neglected ($F_\pi \simeq 87~\MeV$). The matrix field $U$ collects the 
Goldstone bosons of the spontaneously broken chiral symmetry of the QCD 
Lagrangian with three massless flavours, and $D_{\mu}U$ denotes the covariant 
derivative: $D_{\mu}U\!=\!\partial_{\mu}U-ir_{\mu}U+iUl_{\mu}$, in the
presence of external chiral sources $l_{\mu}$ and $r_{\mu}$ of
left-- and right--handed currents. The parameters $g_8$ and $g_{27}$ encode
the dynamics of the integrated-out degrees of freedom in the chiral limit.
These include the heavy quark flavours as well as the light 
hadronic flavour states.
Notice that the octet term proportional to $g_{8}$ induces pure
$\Delta I=1/2$ transitions, while the term proportional to
$g_{27}$ induces both $\Delta I=1/2$ and $\Delta I=3/2$ transitions:
\begin{eqnarray}\label{A0}
A_{0} & = & -\frac{\GF}{\sqrt{2}}V_{ud}V_{us}^{*}\ \sqrt{2}F_{\pi}
\left(g_{8}+\frac{1}{9} g_{27}
\right)(M_{K}^{2}-m_{\pi}^2)\,,
\\ A_{2} & = & -\frac{\GF}{\sqrt{2}}V_{ud}V_{us}^{*}\ 2F_{\pi}
\frac{5}{9}g_{27}(M_{K}^{2}-m_{\pi}^2)\,,
\label{A2}
\end{eqnarray}
where $M_K$ and $m_\pi$ are the masses of the kaon and the pion respectively, 
and $G_F$ is the Fermi four-point interaction parameter.

To calculate $g_8$ and $g_{27}$, we separate the long and short distance 
contributions as usual and perform an Operator Product Expansion (OPE), 
obtaining the effective Hamiltonian 
for $|\Delta S|=1$ transitions~\cite{GLAM,VSZ,GWP},
\begin{equation}
{\cal H}_{ef\hspace{-0.5mm}f}^{\Delta S=1}=\frac{G_F}{\sqrt{2}}
\;\xi_u\sum_{i=1}^8 c_i(\mu)Q_i(\mu)\hspace{1cm}(\mu < m_c=\textrm{charm quark mass})\,,
\label{ham}
\end{equation}
\begin{equation}
c_i(\mu)=z_i(\mu)+\tau y_i(\mu)\;,\hspace*{1cm}\tau=-\xi_t/\xi_u\;,
\hspace*{1cm}\xi_q=V_{qs}^*V_{qd}^{}\;.
\end{equation}
The arbitrary renormalisation scale $\mu$ separates short- and
long-distance contributions to the decay amplitudes. The Wilson 
coefficient functions $c_i(\mu)$ contain all the information 
on heavy-mass scales. For CP conserving processes, the contribution 
involving the CKM elements of the top quark, encoded in $y_i(\mu)$, 
is negligible and only the $z_i(\mu)$ are numerically relevant.
The coefficient functions can be calculated for a scale $\mu \gtrsim 1\,$GeV 
using perturbative renormalisation group techniques. They were computed 
in an extensive next-to-leading logarithm analysis by two groups \cite{BJM,CFMR}. 
After Fierz reordering, the local four-quark 
operators $Q_i(\mu)$ can be written
in terms of color singlet quark bilinears
\begin{eqnarray}
Q_1 &=& 4\,\bar{s}_L\gamma^\mu d_L\,\,\bar{u}_L\gamma_\mu u_L\,, 
\hspace*{2.49cm} 
Q_2 \,\,\,=\,\,\,\,4\,\bar{s}_L\gamma^\mu u_L\,\,\bar{u}_L
\gamma_\mu d_L\,, \label{qia} \nonumber \\[4mm] 
Q_3 &=& \,4\,\sum_q \bar{s}_L\gamma^\mu d_L\,\,\bar{q}_L\gamma_\mu 
q_L\,,  
\hspace*{1.76cm}
Q_4 \,\,\,=\,\,\, \,4\,\sum_q \bar{s}_L\gamma^\mu q_L\,\,\bar{q}_L
\gamma_\mu d_L\,,\nonumber \\[2mm]
Q_5 &=& \,4\,\sum_q \bar{s}_L\gamma^\mu d_L\,\,\bar{q}_R\gamma_\mu 
q_R\,, 
\hspace*{1.74cm}
Q_6 \,\,\,=\,\,\, \,-8\,\sum_q \bar{s}_L q_R\,\,\bar{q}_R d_L\,,
\nonumber \\[1mm]
Q_7 &=& \,4\,\sum_q \frac{3}{2}e_q\,\bar{s}_L\gamma^\mu d_L\,\,
\bar{q}_R \gamma_\mu q_R\,,\hspace*{1.0cm}
Q_8 \,\,\,=\,\,\, \,-8\,\sum_q \frac{3}{2}e_q\,\bar{s}_L q_R\,\,
\bar{q}_R d_L\,, \label{qio}
\end{eqnarray}
where the sum goes over the light flavors ($q=u,d,s$) and
\begin{equation}
q_{R,L}=\frac{1}{2}(1\pm\gamma_5)q\,,\hspace{1cm} 
e_q\,= (2/3,\,-1/3,\,-1/3)\,.
\end{equation}
The operators $Q_3,\ldots,Q_6$ arise from QCD penguin diagrams involving 
a virtual $W$ and a $c$ or $t$ quark, with gluons connecting the virtual heavy 
quark to light quarks. They transform as $(8_L,1_R)$ under $SU(3)_L\times 
SU(3)_R$ and solely contribute to $\Delta I=1/2$ transitions. It is 
important to note that
they are present only below the charm threshold, i.e.~for $\mu< m_c$.
Similarly the Wilson coefficients $z_{7,8}$
of the electroweak penguin operators $Q_{7,8}$ are non-zero 
only for $\mu< m_c$. Thus, in 
the following,
only $Q_1$ and $Q_2$ will be considered as we will always work in the regime
$\mu \gtrsim m_c$ .  Long-distance contributions to the 
amplitudes $A_I$ are contained in the hadronic matrix elements of the four-quark 
operators,
\begin{equation}
\langle Q_i(\mu)\rangle_I \,\equiv\,\langle\pi\pi,\,I|\,Q_i(\mu)\,
|K^0\rangle\,.
\end{equation}

In the strict large $N_c$ limit, i.e.~considering only the W 
exchange diagram with $z_2=1$ we get $g_8=g_{27}=3/5$, while 
experimentally, from Eq.(\ref{gdata}) and Eqs.(\ref{A0})-(\ref{A2}) 
one observes $g^{exp}_8=5.1$ and $g^{exp}_{27}=0.29$. 
This shows how crucial the QCD dynamics is for the $\Delta I = 1/2$ 
rule. Important progress in the understanding of the $\Delta I=1/2$ rule 
was made when it was observed that the short-distance (quark) evolution, 
which is represented by the Wilson coefficient functions in the effective 
Hamiltonian of Eq.(\ref{ham}), leads to both an enhancement of 
the $I=0$ and a suppression of the $I=2$ final state. The {\it octet 
enhancement} \cite{GLAM} in the $(Q_1,Q_2)$ sector is dominated by the 
increase of $z_2$ when $\mu$ evolves from $M_W$ down to $\mu \simeq 1\,$GeV, whereas 
the suppression of the $\Delta I=3/2$ transition results from a partial 
cancellation between the contributions from the $Q_1$ and $Q_2$ operators. 
Taking into account the running of $z_1$ and $z_2$ between $M_W$ and $\mu\simeq m_c=1300$~MeV,
which gives $z_1\simeq-0.5$, $z_2\simeq 1.3$, and still considering 
the matrix elements in the large $N_c$ limit, i.e. considering
only the factorisable contribution, one gets $g_8\simeq 1$ and $g_{27}\simeq 0.5$. 
This gives values closer to the experimental ones but a factor 5 (3/5) is still missing 
for $g_8$ ($g_{27}$). Thus, perturbative QCD effects are far 
from sufficient to describe the $\Delta I=1/2$ rule and QCD dynamics at 
low energies must be addressed beyond the leading $N_c$ limit, that is to say, at the level of the 
non-factorisable contribution \cite{BBG,HKS,BP,BF,Hambye:2003cy}.

Further progress was made when, in addition to the $O(p^2)$ 
weak $\Delta S =1$ Lagrangian of Eq.(\ref{ewcl}), the $O(p^4)$ 
$\Delta S =1$ Lagrangian was also considered. A full fit of all the  
weak Lagrangian constants was then carried out, taking into account not 
only the experimental $K \rightarrow \pi \pi$ amplitudes but also the experimental 
$K \rightarrow \pi \pi \pi$ amplitudes. It was found that the $O(p^2)$ contribution
is expected to account for $g_8=3.3$ and $g_{27}=0.23$. 
The rest of the experimental amplitudes are expected to be explained by 
the $O(p^4)$ $\Delta S = 1$ Lagrangian. Numerically, this $O(p^4)$ higher order (HO) 
contribution is equivalent to adding by hand a contribution of $+1.8$ and $+0.06$ 
to $g_8$ and $g_{27}$, respectively. In the following, we will only calculate the $O(p^2)$ Lagrangian 
constants $g_8$ and $g_{27}$, which should account for about two thirds of Re$A_0$ and 
three quarters of Re$A_2$. Therefore, for comparison with experiment, there are two 
equivalent possibilities: we either compare the values of $g_{8,27}$ we obtain
from Eqs.(\ref{g8int})-(\ref{g27int}) with the values 3.3 and 0.23 above, or, adding 
the $O(p^4)$ contribution by hand, we compare the values we get for 
\begin{eqnarray}
&& g_8^{TOT}=g_8 + g_8^{HO} , \label{HOparts1}  \\
&& g_{27}^{TOT}=g_{27} + g_{27}^{HO} , \label{HOparts2}
\end{eqnarray}
with the values $5.1$ and $0.29$, with $g_8^{HO}=1.8$ and $g_{27}^{HO}=0.06$.

To calculate the non-factorisable contribution to $g_8$ and $g_{27}$, one can 
make use of the chiral symmetry properties of the $\Delta S =1$ effective 
Lagrangian of Eq.(\ref{ewcl}). Instead of calculating the $K \rightarrow \pi\pi$ 
amplitudes explicitly, it is much simpler to calculate them taking $U=1$ 
in Eq.(\ref{ewcl}), i.e. considering the processes with no external pseudoscalar 
and only two external sources coming from the covariant derivatives of $U$.
For reasons explained in Ref.\cite{Hambye:2003cy} it is convenient to consider 
the processes with two external right-handed sources (for instance
for $\cL_8$ we consider $\cL_8 \owns \sum_{i=1,2,3}F_\pi^4 (r_\mu)_{2i} (r^\mu)_{i3} $). 
The non-factorisable contribution to this process of four-quark operators is then 
given by Green's functions involving, on the one hand, the two left-handed 
currents of the four-quark operator inducing this process, and, on the other 
hand, the two right-handed currents coupling to the right-handed sources.
More precisely, including the leading $N_c$ non-factorisable contribution from 
$Q_{1}$ and $Q_2$, the parameters $g_8$ and $g_{27}$ are given by the $Q^2$ integrals 
(with $p$ the Minkowski momentum flowing between the two left-handed currents) 
of the two Green's functions \cite{Hambye:2003cy,PR}:
\begin{eqnarray}
&& {W}_{LRLR}^{\mu\alpha\nu\beta}(p)=\lim_{l\rightarrow 0}\ i^3 \int d^4x d^4y
d^4z\ e^{ipx+il(y-z)} 
\langle 0\vert T\{
L_{\bar{s}d}^{\mu}(x)R_{\bar{d}s}^{\alpha}(y)
L_{\bar{s}d}^{\nu}(0)R_{\bar{d}s}^{\beta}(z)\}
\vert 0\rangle\vert_{\mbox{\rm\tiny conn}} , \label{wlrlr0} \nonumber \\
&& \\
&&{W}_{LLRR}^{\,\mu\,\nu\,\alpha\,\beta}(p) = \lim_{l\rightarrow 0}
\ i^3\int d^4x d^4y d^4z\ e^{ipx + il(y-z)} 
\langle 0\vert T\{L_{\bar{s}u}^{\mu}(x)L_{\bar{u}d}^{\nu}(0)
R_{\bar{d}u}^{\alpha}(y)R_{\bar{u}s}^{\beta}(z)\}\vert
0\rangle\vert_{\mbox{\rm\tiny conn}} , \label{wllrr0} \nonumber \\
&&
\end{eqnarray}
with 
\begin{eqnarray}
&&g_8(\mu) = z_1(\mu) \biggl(-1+\frac{3}{5}g_{\Delta S = 2}(\mu) 
\biggr) + z_2(\mu)\biggl(1-\frac{2}{5}g_{\Delta S = 2}(\mu) - 
\int^{\mu^2}_{0} dQ^2 \frac{{W}_{LLRR}(Q^2)}{4 \pi^2 F_\pi^2}\biggr) , 
\label{g8int} \nonumber \\
&& \\
&&g_{27}(\mu)=\biggl(z_1(\mu)+z_2(\mu)\biggr)
 \frac{3}{5}g_{\Delta S = 2}(\mu)  \,, 
\label{g27int}
\end{eqnarray}
and
\begin{equation}
g_{\Delta S = 2}(\mu)=
1-\frac{1}{32\pi^2 F_\pi^2} \int^{\mu^2}_{0}
dQ^2 \, W_{LRLR}(Q^2), \label{BKint}
\end{equation}
while
\begin{eqnarray}
&&{W}_{LRLR}(Q^2)=
 - \frac{4}{3} \frac{Q^2}{F_\pi^2}\eta_{\alpha\beta}\eta_{\mu\nu} 
\int \frac{d\Omega_{p}}{4 \pi} {W}_{LRLR}^{\mu\alpha\nu\beta}(p) , \label{scalarLRLR} \\
&&{W}_{LLRR}(Q^2)=-\frac{1}{3} \frac{Q^2}{F_\pi^2}
\eta_{\alpha \beta} \eta_{\mu\nu}
\int \frac{d\Omega_{p}}{4 \pi} {W}_{LLRR}^{\,\mu\,\nu\,\alpha\,\beta}(p)\,.
\end{eqnarray}
Notice that we use the notation $Q^2$ for the Euclidean momentum (i.e. $Q^2=-p^2$).

From the above equations we see that the factorisable contributions to $g_8$ and 
$g_{27}$ are 
\begin{eqnarray}
&&g_8^F(\mu) = -\frac{2}{5} z_1(\mu) + \frac{3}{5} z_2(\mu) \label{g8F} ,  \\
&&g_{27}^F(\mu)= \frac{3}{5} \left(z_1(\mu)+z_2(\mu)\right) \label{g27F} . 
\end{eqnarray}
The remaining non-factorisable parts of Eqs.(\ref{g8int})-(\ref{BKint}) are
the subject of the calculations of this paper.

In the above, we only consider, as necessary, the connected 
parts of the four-point functions. 
The currents are defined by 
$R^{\mu}_{{\bar q}_1 q_2}={\bar q}_1 \gamma^\mu \frac{(1+\gamma_5)}{2} q_2$,
$L^{\mu}_{{\bar q}_1 q_2}={\bar q}_1 \gamma^\mu \frac{(1-\gamma_5)}{2} q_2 $, 
and the subscript of $g_{\Delta S = 2}$ comes from the fact that this quantity 
also determines the $\Delta S = 2$ 
transitions (see below).\footnote{Notice that Eq.(\ref{scalarLRLR}) has 
been modified by a normalisation 
factor 4, of which we were not aware in our previous paper \cite{we}.
This normalisation is correct when one sums over all possible 
planar flavour contractions, as we do here.} 

In order to calculate the integrals of Eqs.(\ref{g8int})-(\ref{BKint}), the 
$Q^2$ dependence of $W_{LLRR}$ and $W_{LRLR}$ must be determined.
However, this dependence is known only in the asymptotic 
regimes $Q^2 \rightarrow 0$ and  $Q^2 \rightarrow \infty$.
In the limit when $Q^2 \rightarrow 0$, from $\chi$PT \cite{BP1,PR,Hambye:2003cy}, 
and after a long calculation, one gets
\begin{equation}
W_{LRLR}(Q^2)=6-24({2 l_1+ 5 l_2 + l_3 +l_9})\frac{Q^2}{F_\pi^2}+... ,
\label{chiralLRLR}
\end{equation}
while for the other correlator only one group has calculated the $\chi$PT
result \cite{Hambye:2003cy},
\begin{equation}
W_{LLRR}(Q^2)=-\frac{3}{8}
+(-\frac{15}{2}l_{3}+\frac{3}{2}l_9)\frac{Q^2}{F_{\pi}^2}+...
\label{chiralLLRR}
\end{equation}
In these expressions the $l_i$ are the standard renormalized $O(p^4)$
chiral Lagrangian coefficients, usually denoted by $L_i$. 

In the limit $Q^2 \to \infty$, and using Shifman-Vainshtein-Zakharov OPE 
techniques \cite{VSZ}, one obtains \cite{PR,Hambye:2003cy}
\begin{eqnarray}
\lim_{Q^2\ra\infty} W_{LRLR}(Q^2) & = & +24 
\pi^2\frac{\alpha_s}{\pi}\frac{F_{\pi}^2}{Q^2} , \label{OPELRLR} \\
\lim_{Q^2\ra\infty} W_{LLRR}(Q^2) & = & +\frac{1}{3}\
\pi^2\frac{\alpha_s}{\pi}\frac{F_{\pi}^2}{Q^2}
-\frac{16}{3}\pi^2\frac{\alpha_s}{\pi}
\frac{\langle \bar{\psi} \psi \rangle^2 l_5}{F_{\pi}^4 Q^2} , \label{OPELLRR}
\end{eqnarray}
where $\alpha_s$ is the strong coupling constant, $\langle \bar{\psi} \psi \rangle$ 
is the quark condensate, and $l_5$ is one of the $O(p^4)$ chiral Lagrangian coefficients.
Note that in Eq.(\ref{OPELLRR}) the term depending on $\langle \bar{\psi} \psi \rangle$ 
is numerically dominant.
Using 4D large-$N_c$ diagramatics, one can see that 
$W_{LRLR}$ and $W_{LLRR}$ are given by a sum of simple to triple 
poles in $Q^2$ multiplied by polynomials in $Q^2$. Combining this 
constraint with Eqs.(\ref{chiralLRLR})-(\ref{chiralLLRR}) and 
Eqs.(\ref{OPELRLR})-(\ref{OPELLRR}), we then get the most general form 
for the Green's functions \cite{PR, Hambye:2003cy}
\begin{eqnarray}
W_{LRLR}&=&
\sum_{i=1}^{\infty}\left(\frac{\alpha_i}{(Q^2+M^2_{i})}+\frac{\beta_i}
{(Q^2+M^2_{i})^2}+\frac{\gamma_i}{(Q^2+M^2_{i})^3}\right),\label{generalLRLR}\\
W_{LLRR}&=&
\sum_{i=1}^{\infty}\left(\frac{\alpha'_i}{(Q^2+M^2_{i})}+\frac{\beta'_i}
{(Q^2+M^2_{i})^2}+\frac{\gamma'_i}{(Q^2+M^2_{i})^3}\right) , \label{generalLLRR}
\end{eqnarray}
where the $M_i$'s are the masses of the resonances and $\alpha_i$, 
$\beta_i$, $\gamma_i$, and $\alpha'_i$, 
$\beta'_i$, $\gamma'_i$ are constants.

Quite a few calculations have been proposed in the $1/N_c$ expansion to 
estimate the $Q^2$ integrals \cite{BBG,HKS,BP,Hambye:2003cy}.
They all found a large enhancement of $\mbox{Re}A_0$ together with a 
decrease of $\mbox{Re}A_2$, so that the bulk of the $\Delta I = 1/2$ rule can be 
explained. In those references, the size of the enhancement is determined essentially
by two distinct factors. The first is the $\chi$PT behaviour at low scale, as determined by the 
chiral Lagrangian parameters, and the second is the size of the hadronic scales, 
namely the masses of the hadronic resonances, which will modify and terminate this 
behaviour at some higher scale. In particular, see Ref.\cite{Hambye:2003cy}, which 
explains in detail why the interplay of the relevant hadronic scales---the small 
chiral constants $F_\pi$ and $\langle \bar{\psi} \psi \rangle$ on the factorisable side 
and larger resonance masses on the non-factorisable side---means that such a 
large non-factorisable contribution must be present. However, the effect of 
the resonances was not calculated explicitly in Refs.\cite{BBG,HKS,BP,Hambye:2003cy}, 
but introduced in an indirect way. In \cite{BBG,HKS} only $\chi$PT results 
(Eqs.(\ref{chiralLRLR})-(\ref{chiralLLRR})) were considered, with a cutoff put 
by hand at the mass of the resonances. The work of \cite{Hambye:2003cy} used 
the form of Eqs.(\ref{generalLRLR})-(\ref{generalLLRR}) to interpolate between 
Eqs.(\ref{chiralLRLR})-(\ref{chiralLLRR}) and Eqs.(\ref{OPELRLR})-(\ref{OPELLRR}) 
with a minimum number of resonances, while \cite{BP} employed Nambu-Jona-Lasinio models.
The implicit assumption of these procedures is that the contribution to $W_{LRLR}$ 
and $W_{LLRR}$ from the intermediate momentum region (0.5-2 GeV$^2$) can be obtained 
via a gentle interpolation (i.e.~without ``bumps'') between the chiral behaviour 
and the OPE behaviour. 
In any case, it is clear that a method incorporating 
resonances explicitly would be highly preferable. This is precisely where the power 
of holographic QCD lies, at least for these observables, since it is a method where 
the effect of the entire tower of resonances for each channel can, in principle, 
be calculated.

\subsection{The $B_K$ parameter}

From the Green's function $W_{LRLR}$, there is another observable
whose non-factorisable contribution can be calculated in the 
chiral limit and at leading $N_c$ order, and which can be used as 
a test of the holographic method described below. This is the 
$\hat{B}_K$ observable, parameterising $K^0-\bar{K}^0$ mixing. At 
the quark level, $K^0$ and $\bar{K}^0$ mix due to a box one-loop
diagram where the $K^0$ transforms itself into a $\bar{K}^0$ through 
a pair of W bosons. This diagram leads to the following effective 
Hamiltonian \cite{Nierste}:
\be\lbl{effhal} {\cH}_{\rm eff}^{\Delta
S=2}=\frac{G_{F}^{2}M_{W}^2}{4\pi^2}
\left[\lambda_{c}^2F_1+\lambda_{t}^2F_2+2\lambda_{c}\lambda_{t}F_3\right]
C_{\Delta S=2}(\mu)Q_{\Delta S=2}(x)\, , 
\ee 
with
\be
\label{eq:deltas2} Q_{\Delta S=2}(x)\equiv
(\bar{s}_{L}(x)\gamma^{\mu}d_{L}(x))(\bar{s}_{L}(x)\gamma_{\mu}d_{L}(x)),
\ee 
and $C_{\Delta S = 2}$ is the Wilson coefficient. From this effective 
Hamiltonian, defining
\be
\label{eq:Bpar} <\bar{K}^0|Q_{\Delta S=2}(0)|K^0>\equiv
\frac{4}{3}f_{K}^2M_{K}^2B_{K}(\mu),
\ee
the 
parameter $\hat{B}_K$ is defined as
\be
\hat{B}_K \equiv C_{\Delta S = 2}(\mu) B_K(\mu).
\ee
The large $N_c$ limit (i.e. the factorisable contribution) gives $B_K=3/4$.

In the chiral limit and at leading $N_c$ order, it turns 
out that the non-factorisable contribution is determined by the same integral 
of $W_{LRLR}$ as the one found in parts of $g_8$ and in $g_{27}$, 
Eqs.(\ref{g8int})-(\ref{BKint}). This gives
\be
B_K(\mu)= \frac{3}{4} g_{\Delta S=2}(\mu). 
\ee
These relations come from a dynamical symmetry \cite{diegrundlage} 
relating part of the matrix elements of $Q_1$ and $Q_2$
with those of $Q_{\Delta S=2}$.

Unfortunately, there is no precise experimental determination 
of the $\hat{B}_K$ parameter.
Thus, for our purposes, we will take $\hat{B}_K=0.36 \pm 0.15$ as a reference value, 
as obtained in the chiral limit in Ref.\cite{Hambye:2003cy,PR,Cata:2003mn}. Similar values 
have been obtained in the chiral limit, analytically in Refs.\cite{BKanaly}
and on the lattice in Refs.\cite{BKlattice}. However, note that 
lattice calculations with physical quark masses \cite{BKlatticefull} have been 
shown to be sizeably larger than the chiral limit results, suggesting that the
corrections beyond the chiral limit are large \cite{BKbeyondchiral,PR}.

\section{Analytic Results}

\subsection{Sum of the 5D Witten diagrams}

In this section, we show how to calculate the four-current correlators 
of Eqs.(\ref{wlrlr0}) and (\ref{wllrr0}) in momentum space, i.e.
\begin{eqnarray}
&&  {W}_{LRLR}^{\mu\alpha\nu\beta}(p)= i^3 \lim_{l\rightarrow 0} \langle 0\vert T\{
\tilde L_{\bar{s}d}^{\mu}(p) \tilde R_{\bar{d}s}^{\alpha}(l)
\tilde L_{\bar{s}d}^{\nu}(-p) \tilde R_{\bar{d}s}^{\beta}(-l)\}
\vert 0\rangle  \, , \label{wlrlr0p}  \\
&&  {W}_{LLRR}^{\mu\nu\alpha\beta}(p)= i^3 \lim_{l\rightarrow 0} \langle 0\vert T\{
\tilde L_{\bar{s}u}^{\mu}(p) \tilde L_{\bar{u}d}^{\nu}(-p)
\tilde R_{\bar{d}u}^{\alpha}(l)\tilde R_{\bar{u}s}^{\beta}(-l)\}\vert
0\rangle \, . \label{wllrr0p} 
\end{eqnarray}
These expressions are in momentum space, so we use the tildes to refer to the Fourier 
transformed flavour currents. As we explain above, we use the vector and axial-vector 
field combinations, so that the vector and axial-vector currents are
\begin{eqnarray}
&& \tilde L_{\bar{s}d}^{\mu}(p) = \frac{1}{\sqrt{2}} 
(\tilde J^\mu_{V, \bar{s}d}(p)+\tilde J^\mu_{A, \bar{s}d}(p))
\, , \,\,\,\,\,\,\,\, 
\tilde R_{\bar{s}d}^{\mu}(p) = \frac{1}{\sqrt{2}} 
(\tilde J^\mu_{V, \bar{s}d}(p)-\tilde J^\mu_{A, \bar{s}d}(p)) \, .
\end{eqnarray}
We have the following expansion for ${W}_{LLRR}^{\mu\nu\alpha\beta}(p)$
in terms of the vector and axial-vector currents
\begin{eqnarray}
\frac{i^3}{4}\lim_{l\rightarrow 0} && 
[\tilde J^\mu_{V}(p) \, \tilde J^\nu_{V}(-p)
\tilde J^\alpha_{V}(l) \, \tilde J^\beta_{V}(-l) +
\tilde J^\mu_{V}(p) \, \tilde J^\nu_{V}(-p)
\tilde J^\alpha_{A}(l) \, \tilde J^\beta_{A}(-l) - \nonumber \\
&&
\tilde J^\mu_{A}(p) \, \tilde J^\nu_{V}(-p) 
\tilde J^\alpha_{V}(l) \, \tilde J^\beta_{A}(-l) -
\tilde J^\mu_{A}(p) \, \tilde J^\nu_{V}(-p)
\tilde J^\alpha_{A}(l) \, \tilde J^\beta_{V}(-l) -\nonumber \\
&&
\tilde J^\mu_{V}(p) \, \tilde J^\nu_{A}(-p)
\tilde J^\alpha_{V}(l) \, \tilde J^\beta_{A}(-l) -
\tilde J^\mu_{V}(p) \, \tilde J^\nu_{A}(-p)
\tilde J^\alpha_{A}(l) \, \tilde J^\beta_{V}(-l) +\nonumber \\
&& 
\tilde J^\mu_{A}(p) \, \tilde J^\nu_{A}(-p)
\tilde J^\alpha_{V}(l) \, \tilde J^\beta_{V}(-l) +
\tilde J^\mu_{A}(p) \, \tilde J^\nu_{A}(-p)
\tilde J^\alpha_{A}(l) \, \tilde J^\beta_{A}(-l)] \, ,
\end{eqnarray} 
and similarly for ${W}_{LRLR}^{\mu\alpha\nu\beta}(p)$.

Having calculated in the previous sections the propagators for all the fields in our Lagrangian, 
it is a lengthy but straightforward operation to construct all of the Witten diagrams for 
our four-point functions. One simply uses the bulk-to-boundary propagators 
to connect the four boundary points together through the vertices 
coming from the bulk interaction Lagrangian. Connecting points inside the bulk 
requires a bulk-to-bulk propagator. 

The inclusion of the boundary term Eq.(\ref{boundaryL}) involving $A_5$ obviously increases 
the number of diagrams that contribute to any $n$-point function containing 
the axial-vector current. However, as shown in appendix D, the Ward identities satisfied 
by $W_{LLRR}^{\mu\nu\alpha\beta}$ and $W_{LRLR}^{\mu\alpha\nu\beta}$ can be 
used to demonstrate that one gets the full result by considering purely the diagrams where only
vectors and axial-vectors are connected to the boundary, and where only 
the transverse part of their bulk-to-boundary propagators is taken 
into account. This means that one need not consider the boundary term 
given by $A_\mu\partial_\nu A_5$ for the purpose of this paper.

For both $W_{LLRR}(Q^2)$ and $W_{LRLR}(Q^2)$, the 5D Witten diagram
sum can be split into three distinct classes: diagrams 
where $A_5$ propagates in the bulk, X-diagrams involving the four-boson vertex, 
and Y-diagrams, which involve two three-boson vertices. Each class of diagrams 
contributes to the Green's functions 
at a different order of the momentum $p$: the $A_5$ class contributes
with order $p^0$ and higher, the X-diagrams to order $p^2$ and higher, and the 
Y-diagrams to order $p^4$ and higher. We refer the reader to appendix A for 
an example of each class of diagram, for the
$\tilde J^\mu_{V, \bar{s}d}(p) \, \tilde J^\alpha_{V, \bar{s}d}(l)
\tilde J^\nu_{A, \bar{s}d}(-p) \, \tilde J^\beta_{A, \bar{s}d}(-l)$
contribution to the $W_{LRLR}$ correlator.\footnote{
Note that we must respect the quark-flavour contractions, which eliminates
some of the Witten diagrams. We then draw all the Witten diagrams which 
contribute to each term in this sum, and add all the various parts. 
It turns out that for $W_{LLRR}$ there are 36 distinct diagrams, which gets 
reduced to 24 diagrams upon enforcing the order of quark contraction. 
For $W_{LRLR}$, we find 40 such diagrams which give a non-vanishing 
contribution.  These diagrams are the totality of {\it planar} diagrams
when the order of quark flavour contractions is respected. }

\begin{figure}
\begin{center}
\epsfig{file=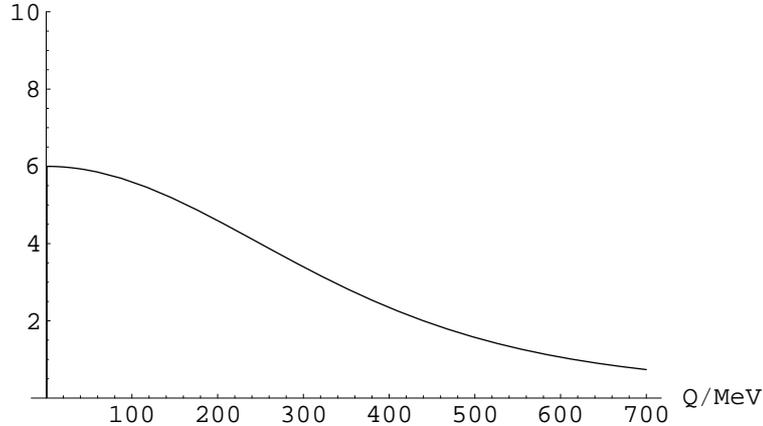, width=10cm} 
\caption{A plot of $W_{LRLR}$ against $Q$, in the limit $L_0\to 0$. 
The correlator $W_{LLRR}$ is $-16$ times 
smaller.} \label{correlator}
\end{center}
\end{figure}

Once we sum the diagrams including all the contributions, we find that 
the two four-point functions are proportional to each other with a factor $-16$.
This factor comes from the $SU(3)$ group theory structure. 
The proportionality is strictly correct only in the $l_\alpha \to 0$ limit, 
which is the limit required for the computation of the $g_8$ and $g_{27}$ parameters. 
We therefore have
\begin{eqnarray}
W_{LRLR}(Q^2) &=& \frac{4i}{3}\frac{Q^2}{F_\pi^2}\Sigma(p=iQ) , \\
W_{LLRR}(Q^2) &=& -\frac{i}{12}\frac{Q^2}{F_\pi^2}\Sigma(p=iQ) , \label{correlators}
\end{eqnarray}
where $\Sigma$ denotes the sum of the diagrams and can be written 
as $\Sigma=\Sigma_{X}+\Sigma_{A_5}+\Sigma_{Y}$, referring to the distinct 
classes of diagrams. The $\Sigma_{X}$ and $\Sigma_{A_5}$ components are given by
\be
\Sigma_{X}(p)=-i\left(\frac{M_5L}{2}\right)\frac{[d-1]^3}{d}
\int\frac{\mathrm{d}z}{z}\left([{K_0^V}^2+{K_0^A}^2]
[{K_p^V}^2+{K_p^A}^2]-4K_0^VK_0^AK_p^VK_p^A\right) ,
\ee
and
\be
\Sigma_{A_5}(p)=-i\left(\frac{M_5L}{\sqrt{2}}\right)^2\frac{[d-1]^2}{d}
\int\frac{\mathrm{d}z}{z}\int\frac{\mathrm{d}z'}{z'}G^5_p(z,z')A'(z,z') ,
\ee
where $d=4$ is the dimension of spacetime and 
\begin{eqnarray}
A'(z,z') &=& 2\left(K_p^A\partial_z K_0^V-K_0^V\partial_z K_p^A\right)
\left(K_0^A\partial_{z'} K_p^V-K_p^V\partial_{z'} K_0^A\right) \nonumber \\
{} &+& \left(K_0^A\partial_z K_p^V-K_p^V\partial_z K_0^A
\right)\left(K_0^A\partial_{z'} K_p^V-K_p^V\partial_{z'} K_0^A\right) \nonumber \\ 
{} &+& \left(K_p^A\partial_z K_0^V-K_0^V\partial_z K_p^A\right)
\left(K_p^A\partial_{z'} K_0^V-K_0^V\partial_{z'} K_p^A\right) .
\end{eqnarray}
As for the Y-diagrams, the integrations are more involved, 
but the sum can be written as $\Sigma_{Y}(p)=Y_1+Y_2+Y_3+Y_4-2Y_5-2Y_6$, where 
\begin{eqnarray}
Y_1 &=&-i\left(\frac{M_5L}{\sqrt{2}}\right)^2[d-1]p^2\int\frac{\mathrm{d}z}{z}
\int\frac{\mathrm{d}z'}{z'}K_p^V(z)G_0^V(z,z')K_p^V(z')K_0^V(z)K_0^V(z') , \nonumber \\
Y_2 &=&-i\left(\frac{M_5L}{\sqrt{2}}\right)^2[d-1]p^2\int\frac{\mathrm{d}z}{z}
\int\frac{\mathrm{d}z'}{z'}K_p^V(z)G_0^A(z,z')K_p^V(z')K_0^A(z)K_0^A(z') , \nonumber \\
Y_3 &=&-i\left(\frac{M_5L}{\sqrt{2}}\right)^2[d-1]p^2\int\frac{\mathrm{d}z}{z}
\int\frac{\mathrm{d}z'}{z'}K_p^A(z)G_0^A(z,z')K_p^A(z')K_0^V(z)K_0^V(z') , \nonumber \\
Y_4 &=&-i\left(\frac{M_5L}{\sqrt{2}}\right)^2[d-1]p^2\int\frac{\mathrm{d}z}{z}
\int\frac{\mathrm{d}z'}{z'}K_p^A(z)G_0^V(z,z')K_p^A(z')K_0^A(z)K_0^A(z') , \nonumber \\
Y_5 &=&-i\left(\frac{M_5L}{\sqrt{2}}\right)^2[d-1]p^2\int\frac{\mathrm{d}z}{z}
\int\frac{\mathrm{d}z'}{z'}K_p^A(z)G_0^A(z,z')K_p^V(z')K_0^V(z)K_0^A(z') ,  \nonumber \\
Y_6 &=&-i\left(\frac{M_5L}{\sqrt{2}}\right)^2[d-1]p^2\int\frac{\mathrm{d}z}{z}
\int\frac{\mathrm{d}z'}{z'}K_p^A(z)G_0^V(z,z')K_p^V(z')K_0^A(z)K_0^V(z') . 
\end{eqnarray}
We perform all the integrals with the limits $L_1$ and $L_0$, and show the full 
results in appendix B. The results are very complicated expressions, but one can 
then take the limit $L_0\to 0$ smoothly. All the divergent contributions cancel, 
and we obtain the simple result
\begin{eqnarray}
\Sigma(Q)_{L_0 \to 0} &=&3iM_5L\left[\frac{16}{Q^6L_1^6}-\frac{14}{5Q^4L_1^4}-\frac{299}{240I_1^2}+
\frac{7}{20Q^2L_1^2I_1^2}+\frac{299}{240I_0^2} \right. \nonumber \\
{} &-&\frac{2}{15Q^2L_1^2I_0^2}+\frac{7}{5Q^4L_1^4I_0^2}
+ \left.\frac{16}{Q^6L_1^6I_0^2}-\frac{32}{Q^6L_1^6I_0}+\frac{13}{12QL_1I_0I_1}\right] , 
\label{sigma} 
\end{eqnarray}
where $I_{0,1}=I_{0,1}(QL_1)$ are the modified Bessel functions of zeroth and first 
order, respectively, and $Q$ is the Euclidean momentum. This simplified form is more 
appropriate for the analysis of the high and low $Q$ behaviour of the correlators.  
See Figure \ref{correlator} for a plot of $W_{LRLR}$ against momentum, 
with $1/L_1=280$ MeV.
Note that $W_{LRLR}$ is found to be positive definite, while $W_{LLRR}$ is 
negative definite because of the proportionality. Both correlators also approach 
zero as $Q \to \infty$, and satisfy the ``sum of poles'' functional form of 
Eqs.(\ref{generalLRLR})-(\ref{generalLLRR}). More precisely, the high $Q$ behaviour of the 
correlators is given by $1/Q^2$, which is the correct functional 
form predicted by perturbative QCD, Eqs.(\ref{OPELRLR})-(\ref{OPELLRR}).

\subsection{The limit $Q\to 0$ and connection with chiral perturbation theory}

\begin{figure}
\begin{center}
\epsfig{file=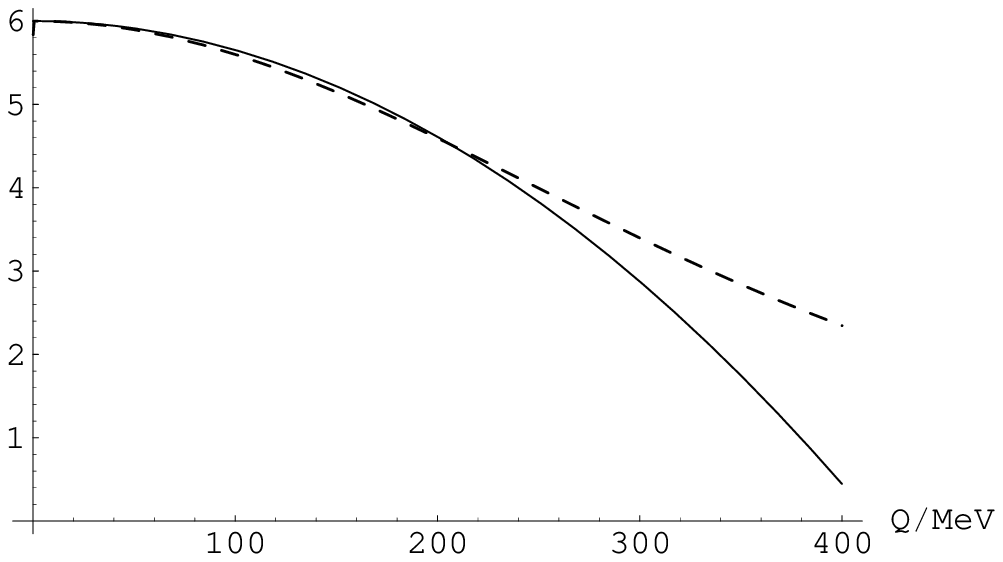, width=10cm} 
\epsfig{file=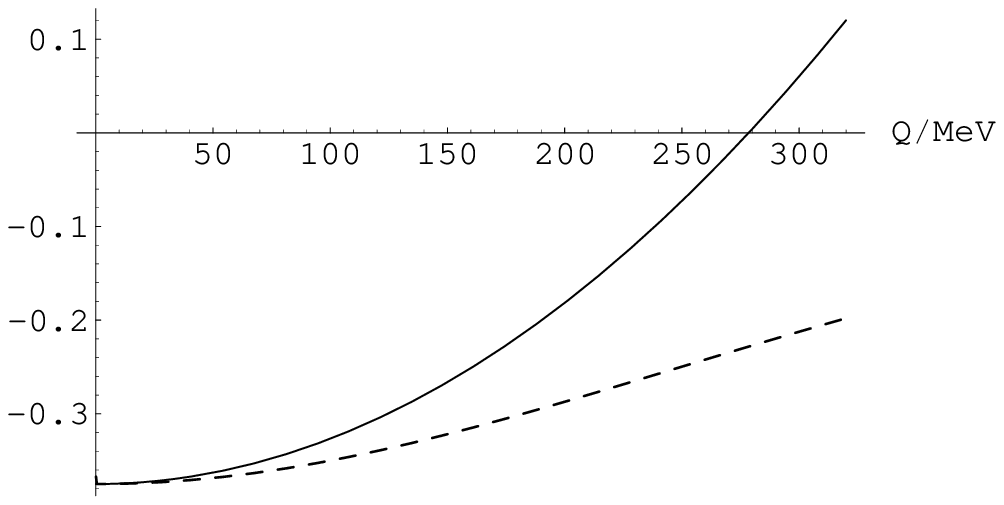, width=10cm} 
\caption{The low $Q$ behaviour of $W_{LRLR}$ (top) and 
$W_{LLRR}$ (bottom) in the $L_0\rightarrow 0$ limit: the 
dashed line is the AdS prediction, 
the solid line is $\chi$PT.} 
\label{LowQplot}
\end{center}
\end{figure}

The pole structure of the propagators for low momentum constitutes a strong 
check on our calculation. Another check is whether our results 
agree with $\chi$PT which, as explained above,  
gives us a constraint on the behaviour of the correlators as $Q\to 0$, 
Eqs.(\ref{chiralLRLR})-(\ref{chiralLLRR}). 
Taking that limit in the expression of $\Sigma(Q)$, we obtain
\be
\lim_{Q^2\ra 0}\Sigma(Q)=3iM_5L\left(\frac{-3}{Q^2L_1^2}
+\frac{105}{64}-\frac{1521}{2560}Q^2L_1^2+\cO(Q^4)\right) \, .
\ee
This is indeed the functional form required by $\chi$PT, 
the $Q^2$ pole being due to the massless pions. Our correlators therefore 
have the low $Q$ behaviour given by
\begin{eqnarray}
\lim_{Q^2\ra 0}W_{LRLR}(Q^2) &=& 6-\frac{105M_5L}{16}\frac{Q^2}{F_\pi^2}
+\cO(Q^4) \, , \label{5dwLRLR}  \\
\lim_{Q^2\ra 0}W_{LLRR}(Q^2) &=& -\frac{3}{8}+\frac{105M_5L}{256}\frac{Q^2}{F_\pi^2}
+\cO(Q^4) \, .\label{5dwLLRR}
\end{eqnarray}
This is to be compared with the expressions obtained via $\chi$PT 
in the chiral limit, Eqs.(\ref{chiralLRLR})-(\ref{chiralLLRR}). 
A plot of our results versus those of $\chi$PT makes things clearer, for 
a value of $1/L_1=280$ MeV (Fig.\ref{LowQplot}). The matching obtained for $W_{LRLR}$ is 
very good for the range of validity of $\chi$PT, while $W_{LLRR}$ does not exhibit as good a 
matching (see below). Note also that the $\chi$PT results shown in the plots 
do not contain any $O(p^6)$ contribution, while our 5D result is to full order in $p$.

In Ref.\cite{Hirn:2005nr}, the coefficients of the 
$\mathcal{O}(p^4)$ chiral Lagrangian were calculated in an AdS setting with identical 
field content to the one used in this work. This allows us to compare our predictions 
in the low momentum limit to those of $\chi$PT with the AdS $l_i$ coefficients calculated 
in Ref.\cite{Hirn:2005nr}. Using those results, and the relations found between the $l_i$ 
coefficients, i.e. $l_2=2 l_1$ and $l_3=-6 l_1$, Eqs.(\ref{5dwLRLR})-(\ref{5dwLLRR}) can be 
rewritten as
\be
\lim_{Q^2\ra 0} W_{LRLR}(Q^2)=6-24({2 l_1+ 5 l_2 + l_3 +l_9})\frac{Q^2}{F_\pi^2}+...
\ee
and 
\be
\lim_{Q^2\ra 0}W_{LLRR}(Q^2)=-\frac{3}{8}+\frac{3}{2}(l_9-l_3)\frac{Q^2}{F_{\pi}^2}+...
\ee
Notice that the first expression coincides exactly
with the pure $\chi$PT calculation, Eq.(\ref{chiralLRLR}).
Similarly, for $W_{LLRR}$, the $O(p^2)$ coefficient $-3/8$ and the $O(p^4)$ $l_9$ 
coefficient $+3/2$ coincide with the corresponding coefficients of Eq.(\ref{chiralLLRR}). 
However, the holographic calculation does not reproduce the  
$O(p^4)$ $l_3$ coefficient of Eq.(\ref{chiralLLRR}), yielding a factor $-3/2$ in place of 
$-15/2$, so that the total $Q^2$ coefficient in $W_{LLRR}$ differs by a factor two approximately 
from the $\chi$PT result. We do not understand this discrepancy. We have performed a variety 
of consistency checks on the 5D calculation, and we do not see any possibility of deviations 
which would alter the proportionality between $W_{LLRR}$ and $W_{LRLR}$. This makes us confident 
that our results are correct.
It seems possible to us that the problem might lie with the sole and rather subtle $\chi$PT 
calculation of the $l_3$ dependence of $W_{LLRR}$. Note also that this difference in 
the $l_3$ coefficient for $W_{LLRR}$ is not significant enough to alter the fact that we 
find a large enhancement for $g_8$ below.

\section{Numerical Results and Discussion}

In this section we present our numerical results for the $\Delta I=1/2$ rule and the $B_K$
mixing parameter. 
To obtain the values of the parameters $g_8$ and $g_{27}$, we must integrate the 
two four-current correlators over the Euclidean momentum as explained in 
Eqs.(\ref{g8int})-(\ref{BKint}). This integral should ideally be regularised 
in the same scheme as the Wilson coefficients of the four-quark operators responsible 
for the kaon decay. These are $z_1$ and $z_2$, and they are usually renormalised 
using dimensional regularisation in three distinct schemes: Leading Order (LO), 
t'Hooft-Veltman (HV) and Naive Dimensional Regularisation (NDR).  On the other hand,
the sharp UV boundary at  $z=L_0$ in the 5D calculation implies that a
hard cut-off at scale $1/L_0$ should be employed in the momentum integral, and that
one should use the appropriate 
values of the Wilson coefficients $z_1$ and  $z_2$ at this energy scale.
We therefore focus upon the Wilson coefficients calculated in the
LO renormalisation scheme as this provides a closer, though admittedly not exact, match
to the holographic part of the calculation.
One may justify this choice as follows: our results for the correlators have the functional 
form expected from QCD calculations in Eqs.(\ref{OPELRLR})-(\ref{OPELLRR}). 
The integrals carried out in the computation of $g_8$ and $g_{27}$ 
will therefore have a logarithmic divergence with respect to the cut-off, rendering the 
integration stable under changes of the high-momentum scale.

We employ two choices of the high-momentum cutoff: 1300 MeV, which is approximately the 
mass of the charm quark, and 1500 MeV.\footnote{In taking the cutoff to be 1500 MeV, 
which is above the charm threshold, we should also consider the contribution of four-quark 
operators involving the charm quark (see. e.g. \cite{Buras}). However, for a scale not larger 
than 1500 MeV, which is well below the mass of the charmed resonances, their contribution 
is expected to be quite small, and in the following we neglect them.} Below, we also 
demonstrate that the results of our calculation are stable against changes in this cutoff.
The values of the Wilson coefficients $z_1$ and  $z_2$ that we use, as calculated 
in the work of \cite{Buras} for the LO scheme, are $z_1=-0.625$, $z_2=1.345$, at 1300 MeV and
$z_1=-0.5699$, $z_2=1.307$,  at 1500 MeV. For $C_{\Delta S = 2}$ we use the values $1.17$ and 
$1.21$, respectively.

Below, we employ a self-consistent prescription to carry out two distinct fits. In the first, 
we fit to the following observables: the mass of the rho meson $m_{\rho}$,
the mass of the $a_1$ axial-vector meson $m_{a_1}$, $F_{\pi}$, $g_8$ and $g_{27}$. 
The values of $F_\pi$, $m_\rho^{exp}$ and $m_{a_1}^{exp}$ that we fit to are 
87 MeV, 776 MeV and 1230 MeV, respectively. For $g_8$ and $g_{27}$, as explained in section 4, 
the values we fit to are $g_8=3.3$ and $g_{27}=0.23$, equivalent 
to $g_8^{TOT}=g_8+g_8^{HO}=5.1$, and $g^{TOT}_{27}=g_{27}+g_{27}^{HO}=0.29$, 
Eqs.(\ref{HOparts1})-(\ref{HOparts2}). 
In the second type of fit, we fit to $\omega$ and $B_K$ instead 
of $g_8$ and $g_{27}$, taking the values
$\omega=1/22.2$ and ${\hat {B}}_K=0.36$, see section 4.
The only free parameters in both fits are $M_5L$ and $L_1$. Inside 
the integral, we use 
our full result for finite $L_0$, the Minkowski version of which 
is shown in appendix B.
The predictions of the model for $F_\pi$, $m_\rho$ and $m_{a_1}$, entering 
in the fits, are 
\cite{Erlich:2005qh,DaRold:2005zs,Hirn:2005nr,we}
\begin{equation}
F_\pi^{th} = \sqrt{\frac{2M_5 L}{L_1^2- L_0^2}} ,  
\label{fpi}
\end{equation}
and, to a good approximation in the range of interest,
\begin{eqnarray}
m_\rho^{th} &\approx & \frac{2.12}{L_1} \frac{(L_1 -0.282 L_0)}{(L_1-L_0)} ,  \\
m_{a_1}^{th} & \approx & \frac{3.38}{L_1} \frac{(L_1 -0.085 L_0)}{(L_1-L_0)} .
\label{mrho_ma1}
\end{eqnarray}

\begin{center}
\begin{table}[h]
\begin{center}
\begin{tabular}{|c|c|c|c|c|}\hline
\multicolumn{1}{|c}{Observable}&
\multicolumn{1}{|c|}{A}&
\multicolumn{1}{c|}{B}&
\multicolumn{1}{c|}{C}& 
\multicolumn{1}{c|}{D} 
\\ \hline \hline
$L_0^{-1}$                    & $1300$ MeV & $1500$ MeV & $1300$ MeV & $1500$ MeV \\
$L_1^{-1}$                    & $274$  MeV & $275$  MeV & $277$  MeV & $280$  MeV \\
$m_\rho^{th}/m_\rho^{exp}$    & $0.91$     & $0.90$     & $0.93$     & $0.92$     \\
$m_{a_1}^{th}/m_{a_1}^{exp}$  & $0.95$     & $0.93$     & $0.97$     & $0.95$     \\
$F_\pi^{th}/F_\pi$            & $1.15$     & $1.17$     & $1.12$     & $1.14$     \\
$g_{8}^{TOT}/g_{8}^{exp}$     & $0.74$     & $0.72$     & $0.75$   & $0.74$   \\
$g_{27}^{TOT}/g_{27}^{exp}$   & $0.85$     & $0.85$     & $0.79$   & $0.78$   \\
$1/\omega$                    & $19.5$   & $19.2$   & $21.4$     & $21.3$     \\
$\hat{B}_K^{th}$              & $0.38$   & $0.38$   & $0.34$     & $0.34$     \\
\hline
\end{tabular}
\end{center}
\caption{Columns $A, B$ show a fitting to $m_{\rho}$, $m_{a_1}$, $F_{\pi}$, $g_8$ and $g_{27}$.
Columns $C, D$ show a fitting to $m_{\rho}$, $m_{a_1}$, $F_{\pi}$, $1/\omega$ and $\hat{B}_K$.
Note that, as explained in section 4, $g_{8}^{TOT}=g_8+1.8$, $g_{27}^{TOT}=g_{27}+0.06$, 
where $g_8$ and $g_{27}$ are the quantities we calculated from the AdS model.
We use the values $F_{\pi}=87$ MeV, $m_{\rho}^{exp}=776$ MeV, $m_{a_1}^{exp}=1230$ MeV,
$g_{8}^{exp}=5.1$ and $g_{27}^{exp}=0.29$.}
\end{table}
\end{center}

The results for both fits are given in Table 1 and are quite similar. 
Taking into account the relative crudeness of our model, and the use of 
the large $N_c$ expansion of QCD, we find these results quite good.  
The discrepancy with experiment never exceeds $\sim 25 \%$,
and for some observables is much smaller.  It must be emphasised again 
that, having picked the values for the upper cut-off 
to be 1300 MeV and 1500 MeV and thereby fixed $L_0$, the only remaining free 
parameters are $L_1$ and $M_5L$.  Therefore, we fit five independent 
observables with only two free parameters.

To understand the structure of the $g_8$ and $g_{27}$ results in Table~1, 
it is useful to decompose the numerical results into three components: the leading $N_c$ factorised 
part of Eqs.(\ref{g8int})-(\ref{BKint}) given explicitly in Eqs.(\ref{g8F})-(\ref{g27F}), 
the $1/N_c$ non-factorisable contribution of Eqs.(\ref{g8int})-(\ref{BKint}) 
(encoded in the $W_{LRLR}$ and $W_{LLRR}$ Green's functions we calculated), and the contribution
from the higher-order corrections, $g_8^{HO}$ and $g_{27}^{HO}$, which as said above we take from 
Ref.\cite{KMW}. For example, for the results of column A this decomposition goes as follows:
$g_8^{TOT}= 1.06+0.92+1.8=3.78$ and $g_{27}^{TOT}=0.43-0.24+0.06=0.25$, whereas the 
experimental values to compare with are $g_8^{exp}=5.1$ and $g_{27}^{exp}=0.29$, respectively. We 
see that, for $g_8$, the non-factorisable chiral limit contribution of $0.92$ 
is of the same order as the factorised contribution of $1.06$, and effectively doubles it. 
For $g_{27}$, the non-factorisable contribution of $-0.24$ effectively 
divides the factorised result of $0.43$ by more than two. As already mentioned above, 
the fact that the $1/N_c$ contribution can be as large as the factorised part, even 
though the $1/N_c$ series is still expected to converge, can be naturally 
explained by the interplay of the various hadronic scales involved in 
$g_8$ and $g_{27}$ \cite{Hambye:2003cy}.   

For a rather conservative estimate of the error\footnote{This is also useful as an estimate of 
higher order $1/N_c$ correction effects.} involved in our calculation,
it is interesting to compare the
results of Table~1 with the values we obtain by fitting only $F_{\pi}$, $m_\rho$ 
and $m_{a_1}$. In this case, for example with $L_0^{-1}=1300$~MeV, we 
get $F_{\pi}^{th}/F_{\pi}^{exp}=1.00$, $m_{\rho}^{th}/m_{\rho}^{exp}=0.98$, 
$m_{a_1}^{th}/m_{a_1}^{exp}=1.02$, and $g_8^{TOT}=1.06+1.32+1.8=4.18$, which gives 
$g_8^{TOT}/g_8^{exp}=0.82$.
In this case, the non-factorisable chiral limit contribution of 1.32, is $\sim 40 \%$ larger 
than the value 0.92 above. For $g_{27}$ we get
a smaller result: $g_{27}^{TOT}=0.43-0.35+0.06=0.14$, so that the non-factorisable 
chiral limit contribution of $-0.35$, is 
also $\sim 40 \%$ larger than the value $-0.24$ above. 
The small total result it gives is not surprising, as $g_{27}$ involves
the difference of two large positive 
contributions, i.e. a $1/N_c^2$ correction of order $\sim 30 \%$ of 
our $1/N_c$ contribution would bring $g_{27}$ close 
to the experimental value.

Note that our results are similar to what has been obtained in other 
analytical calculations utilising the $1/N_c$ expansion \cite{HKS,Hambye:2003cy,BP}. 
In particular, in Ref.\cite{Hambye:2003cy} the values $g_{8}^{TOT}/g_{8}^{exp}=0.76$
and $g_{27}^{TOT}/g_{27}^{exp}=0.79$ have been obtained.
The two methods are, however, quite different, as explained in section 4. The main 
advantage of our model is that it allows the calculation of the four-point 
functions in the entire relevant momentum region within one consistent 
setting, thereby removing the need for interpolation in any specific 
momentum range. 

One could ask why we took the parameter $M_5 L$ 
as a free parameter in the fits. 
In Refs.\cite{Son:2003et,Erlich:2005qh,DaRold:2005zs,Hirn:2005nr,DaRold:2005vr,Katz:2005ir,we}, 
it has been shown that this parameter determines 
the high $Q^2$ logarithm of the vector-current two-point correlator, 
so that it can be fixed from a matching with the corresponding QCD 
coefficient \cite{Shifman:1978bx}, yielding
\be
M_5 L = \frac{N_c}{12 \pi^2} \, .
\ee
The point is that, for two-point functions only, there is enough parameter freedom 
to match the QCD logarithm (see in particular \cite{Erlich:2005qh,DaRold:2005zs,Hirn:2005nr}),
unlike the more complicated case presented here. Clearly, it is not expected that the model we consider, 
based on a simple slice of AdS with a hard cut-off in the UV at the scale $L_0^{-1}$, would lead to 
the exactly correct QCD behaviour. However, comparing our results with those of QCD  
in Eqs.(\ref{OPELRLR})-(\ref{OPELLRR}), we observe that for both $W_{LRLR}$ and $W_{LLRR}$ 
we reproduce the good functional behaviour (i.e. the $1/Q^2$ dependence). 
We also observe that we get the correct sign for the coeffcients of $1/Q^2$ and, amazingly, 
we even get values for these coefficients which are within a factor $2-3$ of the perturbative 
QCD result, for $Q^2\sim 2-5$ GeV$^2$. Note that this is a very surprising outcome, especially 
for $W_{LLRR}$, because the high-momentum dependence of the latter involves the 
quark condensate contribution, as seen in Eq.(\ref{OPELLRR}). In fact, the dominance 
of the quark condensate term in Eq.(\ref{OPELLRR}) guarantees that $W_{LLRR}$ is negative 
in the far UV, and so we predict its sign correctly, although our model has no 
equivalent of the quark condensate (or of $\alpha_s$ for that matter). As mentioned above, 
one way of introducing a tunable condensate is by adding a scalar field in the bulk. It would 
be interesting to see whether the calculation of $W_{LLRR}$ in that case gives a more accurate 
description of the high-momentum behaviour. 

We showed in section \ref{totalL} that the 5D bulk Lagrangian we employ is the leading 
order Lagrangian in the large-$N_c$ expansion, and that operators of higher mass dimension 
are sub-leading in $N_c$. In principle, these operators may contribute 
to the four-point functions calculated here. In this section, we have presented the results of a 
fit of five observables using only two independent parameters, which 
are the IR brane position $L_1$ and the dimensionless combination $M_5L$. Thus, the fact that 
a fit using only the leading operator of Eq.(\ref{Lagrangian}) gives good agreement to the data is 
non-trivial. If we were to introduce the full set of sub-leading operators that contribute, 
then we increase the number of free parameters of the model (because the coefficients
of the new operators are unconstrained by bulk gauge invariance or 
other symmetries) and the fit loses predictivity. Therefore, the success of the 
restricted fit performed here shows that the coefficients of the sub-leading 
operators are not anomalously large.

A technical, but important, issue to mention concerns the gauge symmetry 
of the 5D theory. To carry out the 5D calculations, we had to choose a specific gauge 
to work in, and we picked the convenient $R_\xi$ gauge taken in the limit $\xi \to \infty$. Now,
it is clear that the results of the holographic calculation must be independent of the gauge parameter $\xi$, and 
of any choice of gauge-fixing.  This must, in fact, be a feature of any holographic 
calculation involving gauge freedom in the AdS theory. In previous AdS/QCD computations 
there had been no need for concern, since these calculations only involved two-point 
functions and all the propagators in the 5D theory were boundary-to-boundary ones. Boundary 
theory current conservation then automatically projects out the longitudinal gauge-dependent 
component of the 5D propagators, rendering the results gauge-independent. 
Unfortunately, the situation is less clear in the case of four-current correlators, because 
the latter involve Witten diagrams with explicitly gauge-dependent propagators 
(exchange diagrams \cite{Freedman}). In the above, we trust that the power of AdS/CFT 
guarantees that any such holographic calculation will yield results that are independent 
of the 5D gauge fixing. 

The model has many shortcomings, due to its crude nature. For example, a concern for this class of 
models is the behaviour of the masses of the resonances $M_{V_n}$ and $M_{A_n}$ as $n$ approaches 
infinity. One finds that the simple treatment presented here shows that the masses 
of the Kaluza-Klein modes go like $n$ for large $n$, in sharp contrast to the predictions 
of large-$N_c$ theories, in which the masses of the resonances go like $\sqrt{n}$. A 
recent paper \cite{Karch:2006pv} has shown, however, that with a more sophisticated 
handling of the IR truncation of the AdS space, one can indeed recover the large-$N_c$ 
Regge behaviour.  Whether this will improve the results obtained here remains to be 
seen. Finally, one must keep in mind that the calculations done here were all in 
the chiral limit. In order to account for massive quarks, one would have to introduce the 
bi-fundamental scalar of Refs.\cite{Erlich:2005qh,DaRold:2005zs}. In particular, this 
would allow the inclusion of the mass of the strange quark, whose effect could easily 
be as large as 20-25$\%$ for $g_8$, or even more for $g_{27}$, due to the cancellation 
of factorizable and non-factorizable contributions in the latter.

\section{Conclusions}
\lbl{conclusions}

In this paper we have calculated, within the simplest possible version 
of holographic QCD, the four-point flavour current correlators crucial to the resolution 
of the $\Delta I = 1/2$ puzzle of QCD. We believe that our results are quite 
encouraging for the AdS/QCD approach. The holographic theory automatically and consistently 
includes the contributions of the infinite tower of meson resonances to the four-point 
correlators. We also reproduce, to a good level of accuracy, the low-momentum 
and high-momentum behaviour of these correlators, as deduced from chiral perturbation 
theory and perturbative QCD, respectively. This agreement is particularly impressive 
for the correlator $W_{LRLR}$. Moreover, the results of a fit of the holographic predictions 
to the experimental data agree well, with $25 \%$ accuracy or better, showing that the 
dynamics of the $\Delta = 1/2$ rule is operative in AdS/QCD. 
For quantities as difficult to calculate as the isospin amplitudes of kaon decay 
Re$A_0$ and Re$A_2$, this is remarkable.

A rather obvious limitation of the model concerns the description of $\chi$SB. As explained 
above, although the imposition of IR boundary conditions on the bulk $SU(3)_L \times SU(3)_R$ 
gauge fields correctly incorporates the leading $\chi$SB behaviour, a bi-fundamental bulk scalar is needed 
to fully account for the physics of $\chi$SB. The inclusion of this 
field will directly introduce pseudo-scalar resonances into the 4D field content, 
and these will indeed have relevant contributions to the four-current correlators 
calculated here. We will also have an extra parameter that can be tuned 
\cite{Erlich:2005qh,DaRold:2005zs}, corresponding to the quark condensate.
We have also not included the effects of the anomalous $U(1)_A$ symmetry of QCD,
nor the explicit breaking of chiral symmetry due to bare quark masses.  One  
envisions these improvements having a complicated yet positive effect on the calculation of 
four-point current correlators presented in this paper.    

We believe that the results of this paper for four-point functions show that 
it is worth investigating further the predictions of AdS/QCD.

~

\centerline{\bf Acknowledgements}

~

This work has been supported by PPARC Grant PP/D00036X/1, and  
by the `Quest for Unification' network, MRTN-CT-2004-503369. The work
of B.H. was supported by the Clarendon Fund, Balliol College,
and Christ Church College. The work of T.H. 
was supported by a Ram\'on y Cajal contract of the Ministerio de 
Educaci\'on y Ciencia of Spain.
T.H., B.H. and J.M.R acknowledge the hospitality of the CERN Theory Group
where part of this work was carried out. J.M.R also acknowledges the 
Galileo Galilei Institute, Florence for
hospitality, and the INFN for partial support. M.S. also acknowledges the
ECT$^*$, Trento for hospitality.
We thank Eduardo De Rafael, Nick Evans, Johannes Hirn, Santiago Peris 
and Ver\'onica Sanz for discussions and correspondence.

\newpage

\appendix
\appendixpage
\addappheadtotoc

\section{The different classes of 5D Witten diagrams}

We adopt the following conventions in labeling the vector, axial-vector and pseudoscalar
propagators:
\begin{center}
\epsfig{file=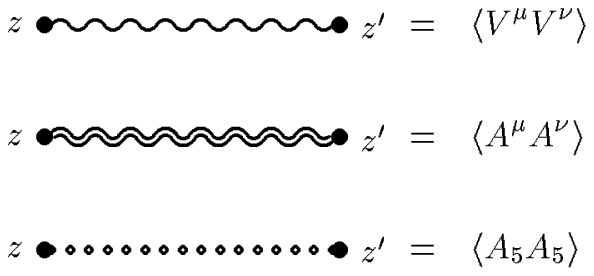, width=6cm, height=3cm}
\end{center}
Using the vertices calculated from the full Lagrangian, we can then draw all the Feynman 
diagrams that contribute to a specific four-point function in momentum space. To illustrate
this, we consider here the $<J_V J_V J_A J_A>$ piece contributing to $W_{LRLR}$. This
is given by the following sum of diagrams:
\begin{center}
\epsfig{file=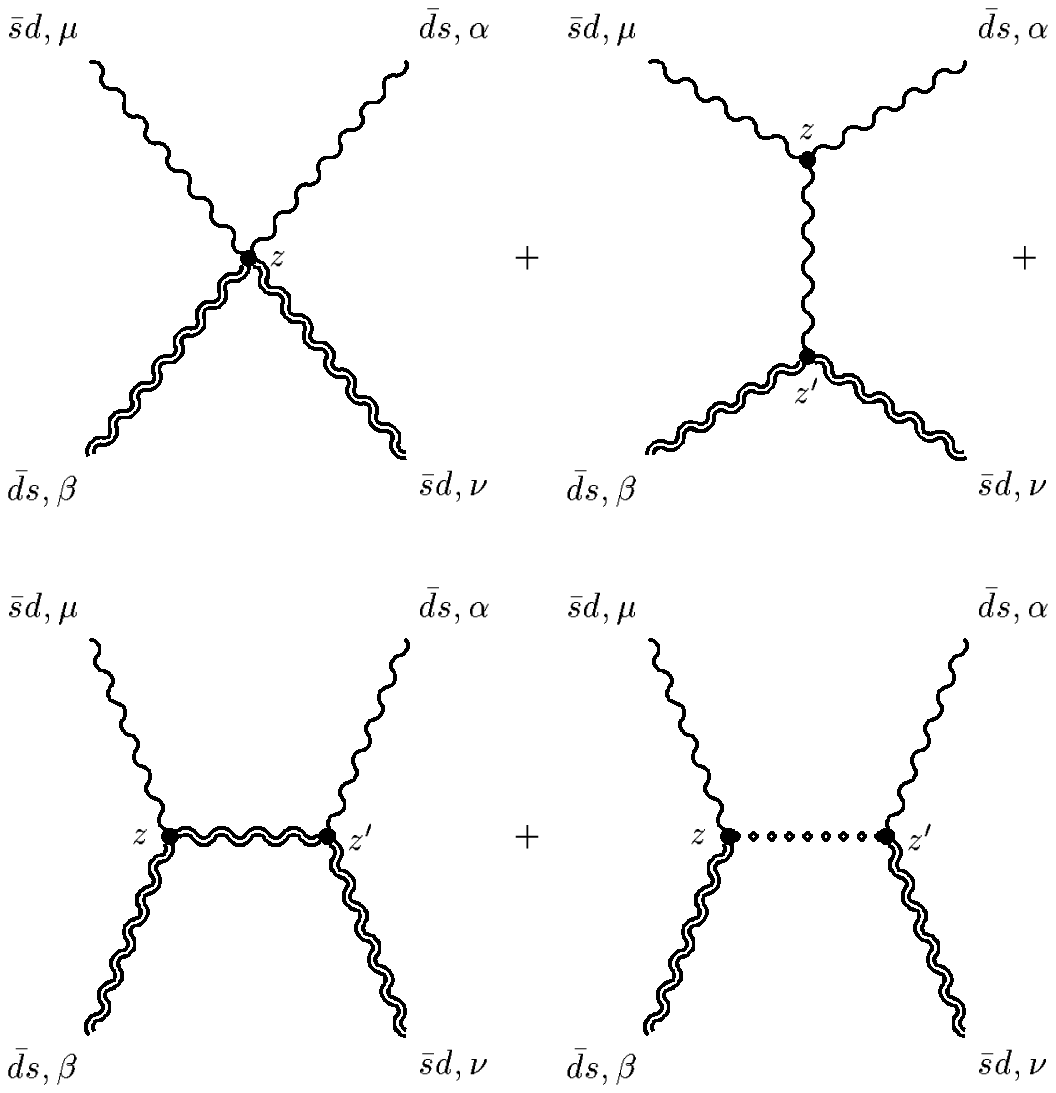, width=11cm, height=11cm}
\end{center}
In principle, one could have two more ``cross diagrams'', but they are eliminated by 
the large $N_c$ condition that diagrams must be planar.

\newpage

\section{The sum of Witten diagrams for finite $L_0$}

Here we present the results of the Witten diagram summation for finite $L_0$ 
in Minkowski space. As in section 5, we write $\Sigma=\Sigma_{X}+\Sigma_{A_5}+\Sigma_{Y}$.
Defining
\begin{eqnarray}
C_n(z)&=&C\mathcal{J}_n(pz)+D\mathcal{Y}_n(pz)  , \nonumber \\
\cc_n(z)&=&\cc \mathcal{J}_n(pz)+\cd \mathcal{Y}_n(pz)  , \nonumber \\
F(z)&=&p^2[z^2C_2(z)+L_1^2C_0(z)] , \nonumber \\
G(z)&=&p^4z^4C_2(z)-2p^3z^3C_3(z)-p^4L_1^2z^2C_2(z) . \nonumber
\end{eqnarray}
we obtain
\begin{equation}
\Sigma_{A_5}=\frac{36iM_5L}{L_0^2(L_1^2-L_0^2)^3(AD-BC)^2p^4}
\left( \left[ z^2C_2(z) \right]\Big\vert_{L_0}^{L_1} \right)^2 , \nonumber
\end{equation}
and 
\begin{eqnarray}
\Sigma_{X}&=&3iM_5\left(\frac{9}{8}\right)\Bigg[ \frac{1}{L_0^2(AD-BC)^2}\left( \frac{z^2}{2}(C_1(z)^2-C_0(z)C_2(z)) 
\right)  \nonumber \\
&+& \frac{1}{L_0^2(A\cd -B\cc )^2}\left( \frac{z^2}{2}(\cc_1(z)^2-\cc_0(z)\cc_2(z)) \right) \nonumber \\
&+& \frac{1}{L_0^2(L_1^2-L_0^2)^2(AD-BC)^2}\Bigg( \frac{-z^4(\frac{6}{5p^2}+L_1^2)}{3}(C_1(z)^2+C_2(z)^2) 
\nonumber \\
&+& \frac{L_1^4z^2}{2}(C_1(z)^2-C_0(z)C_2(z))+\frac{z^4}{10p^2}\left((p^2z^2+3)C_1(z)^2
+\frac{p^2z^2}{4}(C_0(z)+3C_2(z))^2 \right) \Bigg) \nonumber \\
&+& \frac{1}{L_0^2(L_1^2-L_0^2)^2(A\cd -B\cc )^2}\Bigg( \frac{-z^4(\frac{6}{5p^2}+L_1^2)}{3}
(\cc_1(z)^2+\cc_2(z)^2) \nonumber \\
&+& \frac{L_1^4z^2}{2}(\cc_1(z)^2-\cc_0(z)\cc_2(z))+\frac{z^4}{10p^2}\left((p^2z^2+3)\cc_1(z)^2
+\frac{p^2z^2}{4}(\cc_0(z)+3\cc_2(z))^2 \right) \Bigg) \nonumber \\
&+& \frac{4}{L_0^2(L_1^2-L_0^2)(AD-BC)(A\cd -B\cc )}\left( \frac{z^4}{6}C_2(z)\cc_2(z) \right. \nonumber \\
&+& \left. \left(\frac{z^4}{6}-\frac{L_1^2z^2}{2} \right)C_1(z)\cc_1(z)
+\frac{L_1^2z^2}{4}(C_0(z)\cc_2(z)+C_2(z)\cc_0(z)) \right) \Bigg] _{L_0}^{L_1} . \nonumber
\end{eqnarray}
Writing $\Sigma_{Y}(p)=Y_1+Y_2+Y_3+Y_4-2Y_5-2Y_6$, we obtain
\begin{eqnarray}
Y_1&=&\frac{3iM_5L}{4L_0^2(AD-BC)^2}\left(\Big[z^2C_1(z)^2 \Big]_{L_0}^{L_1}-L_0^2C_0(L_0)^2 \right) , \nonumber \\
Y_2&=&\frac{3iM_5Lp^2}{4L_0^2(L_0^2-L_1^2)^3(AD-BC)^2}
\Bigg(\frac{L_1^2-L_0^2}{p^8}\left[\frac{p^6z^6}{5}[C_1(z)C_3(z)+C_2(z)C_4(z)] \right.   \nonumber \\
&+& \frac{p^6z^4L_1^2}{3}[C_1(z)^2+C_2(z)^2]-L_1^4p^6z^2[C_1(z)^2-C_0(z)C_2(z)] \nonumber \\
&-& \left. 2p^2L_1^2\left(C_0(z)[p^3z^3C_1(z)-4p^2z^2C_2(z)]
+\frac{p^4z^4}{6}[C_0(z)C_2(z)+C_1(z)C_3(z)] \right) \right]_{L_0}^{L_1}  \nonumber \\
&+& \frac{1}{p^8}\left[(L_0^2F(L_0)-L_1^2F(L_1))(G(L_1)-G(L_0))-(L_0^2G(L_1)-L_1^2G(L_0))(F(L_1)-F(L_0)) \right] \nonumber \\
&+& \frac{1}{p^{10}}(G(L_1)-G(L_0))^2+\frac{L_0^2L_1^2}{p^6}(F(L_1)-F(L_0))^2 \Bigg) , \nonumber \\
Y_3&=&\frac{3iM_5L}{4L_0^2(L_0^2-L_1^2)(A\cd -B\cc )^2p^4}\Big[ p^4L_1^2(L_1^2-L_0^2)\cc_0(L_1)\cc_2(L_1) \nonumber \\
&+& p^4L_0^2(L_1^2-L_0^2)[\cc_0(L_0)^2+\cc_1(L_0)^2]+4p^2L_0^2\cc_1(L_0)^2 \Big]  , \nonumber \\
Y_4&=&\frac{3iM_5Lp^2}{4L_0^2(L_1^2-L_0^2)^2(A\cd -B\cc )^2}\Bigg( \left[ -\frac{z^6}{5p^2}[\cc_1(z)\cc_3(z)+\cc_2(z)\cc_4(z)] 
\right. \nonumber \\
&+&  \left.  \frac{L_1^2z^4}{3p^2}[\cc_1(z)(\cc_3(z)-\cc_1(z))+\cc_2(z)(\cc_0(z)-\cc_2(z))] \right]_{L_0}^{L_1}
-\frac{L_1^6}{p^2}\cc_0(L_1)\cc_2(L_1) \nonumber \\
&-& 2\frac{L_1^5}{p^3}\cc_0(L_1)\cc_3(L_1)-\frac{L_1^4L_0^2}{p^2}[\cc_1(L_0)^2+\cc_0(L_0)^2]  
-2\frac{L_0^3}{p^3}\cc_2(L_0)[L_0^2\cc_3(L_0)+L_1^2\cc_1(L_0)] \Bigg) , \nonumber \\
Y_5&=& \frac{3iM_5Lp^2}{4L_0^2(L_1^2-L_0^2)^2(A\cd -B\cc )(AD-BC)}
\left[2\frac{(L_0^2-L_1^2)}{p^2}\left( \frac{z^4}{6}[\cc_1(z)C_1(z)+\cc_2(z)C_2(z)] \right. \right. \nonumber \\
&-& \left. \left.  \frac{z^2L_1^2}{4}[2\cc_1(z)C_1(z)-\cc_2(z)C_0(z)-\cc_0(z)C_2(z)] \right)
-2\frac{L_0\cc_1(L_0)}{p^7}[G(z)-p^2L_1^2F(z)] \right]_{L_0}^{L_1} , \nonumber \\
Y_6&=& \frac{3iM_5Lp^2}{4L_0^2(L_0^2-L_1^2)(A\cd -B\cc )(AD-BC)}
\Bigg( \frac{2L_0C_0(L_0)}{p^3}[L_0^2\cc_3(L_0)+L_1^2\cc_1(L_0)] \nonumber \\
&+& \frac{1}{p^6}\left[\frac{p^4L_1^4}{3}C_2(L_1)\cc_2(L_1)+2p^3L_1^3C_1(L_1)\cc_0(L_1)
+\frac{1}{2}(8+p^2L_1^2)p^2L_1^2C_2(L_1)\cc_0(L_1) \right. \nonumber \\
&+& \frac{1}{2}(8+p^2L_1^2)p^2L_0^2[2C_1(L_0)\cc_1(L_0)-C_0(L_0)\cc_2(L_0)-C_2(L_0)\cc_0(L_0)] \nonumber \\
&-& \left. \frac{p^4L_0^4}{3}[C_1(L_0)\cc_1(L_0)+C_2(L_0)\cc_2(L_0)]
-2p^3L_0^3[C_1(L_0)\cc_0(L_0)+C_2(L_0)\cc_1(L_0)] \right] \Bigg) . \nonumber
\end{eqnarray}

\newpage

\section{The axial two-point function}

The axial two-point function is defined by
\begin{equation}
<J_A^\mu(x)J_A^\nu(x')>= -i \int \mathrm{d}^4p \, e^{-ip(x-x')} \, \Pi_A^{\mu\nu}(p) . \nonumber
\end{equation}
The Ward identities require a transversal $\Pi_A^{\mu\nu}(p)$, so that we may write 
$\Pi_A^{\mu\nu}(p)=\Pi_A(p^2)T^{\mu\nu}$, where $T^{\mu\nu}$ is the transversal projector 
in $p$. 
To calculate this two-point function in our AdS/QCD setup, we have to use the boundary 
Lagrangian as shown in Eq.(\ref{boundaryL}). This
allows us to write an expression for the axial current by differentiating the full 
Lagrangian with respect to the boundary source $a_\mu$. Schematically, we obtain the 
following expression
\begin{equation}
J_A^\mu(x)=\frac{M_5L}{z}[-\partial_zA_\mu(x,z)+\partial_\mu A_5(x,z)]\Big\vert_{L_0} . \nonumber
\end{equation}
Plugging this expression into the equation for the axial two-point function, we find that 
we can write the result as the sum of the propagators of the $A_\mu$ and $A_5$ fields, 
giving
\begin{eqnarray}
<J_A^\mu(x)J_A^\nu(x')>&=&\frac{M_5L}{z'}\partial_{z'}\left(\frac{M_5L}{z}\partial_z
<A_\mu(x,z)A_\nu(x'z')>\right)\Big\vert_{z,z'=L_0} \nonumber \\
&+&\left(\frac{M_5L}{L_0}\right)^2\partial_{\mu}^{'} \partial_\nu<A_5(x,L_0)A_5(x',L_0)>. \nonumber
\end{eqnarray} 
Now, making use of the definitions of the axial propagators found in section 5 above, 
it is easy to see that the $A_5$ propagator cancels the longitudinal part of the $A_\mu$ 
bulk-to-boundary propagator. Thus, the boundary term containing $A_5$ is essential for 
the satisfaction of the transversality condition. 

\newpage

\section{Simplification to transverse boundary propagators}
 
The proof that only the transverse part of the boundary propagators is necessary for the
calculation of $W_{LLRR}(Q^2)$ and $W_{LRLR}(Q^2)$
is as follows: Consider a general four-point function with external momenta $p_\mu$, $p_\nu$, $l_\alpha$ 
and $l_\beta$. We assume the most general boundary propagators, i.e. with transverse and 
longitudinal parts. We can write this general four-point function as
\begin{eqnarray}
W_4^{\mu\nu\alpha\beta} &=& c_1 T^\alpha T^\beta T^\mu T^\nu \nonumber \\ 
&+& d_1 T^\alpha T^\beta T^\mu p^\nu + d_2 T^\alpha T^\beta p^\mu T^\nu 
+ d_3 T^\alpha l^\beta T^\mu T^\nu+d_4 l^\alpha T^\beta T^\mu T^\nu \nonumber \\
&+& e_1 T^\alpha T^\beta p^\mu p^\nu + e_2 l^\alpha l^\beta T^\mu T^\nu +
e_3 T^\alpha l^\beta T^\mu p^\nu + e_4 T^\alpha l^\beta p^\mu T^\nu \nonumber \\
&+& e_5 l^\alpha T^\beta p^\mu T^\nu + e_6 l^\alpha T^\beta T^\mu p^\nu \nonumber \\
&+& f_1 T^\alpha l^\beta p^\mu p^\nu + f_2 l^\alpha T^\beta p^\mu p^\nu 
+f_3 l^\alpha l^\beta T^\mu p^\nu + f_4 l^\alpha l^\beta p^\mu T^\nu \nonumber \\
&+& g_1 l^\alpha l^\beta p^\mu p^\nu . \nonumber 
\end{eqnarray} 
In this expression, $T^\mu$ and $T^\nu$ are transverse projectors in $p^\mu$ and 
$p^\nu$ respectively, whereas $T^\alpha$ and $T^\beta$ are transverse projectors 
in $l^\alpha$ and $l^\beta$. We omit the second index of every transverse projector, since  
it is contracted inside the coefficient functions $c_i \to g_i$, where all the vertex 
and gauge structure resides. For instance, $d_1 T^\alpha T^\beta T^\mu p^\nu$ is really 
$d_{1, \alpha' \beta' \mu' \nu'} T^{\alpha\alpha'} T^{\beta\beta'} T^{\mu\mu'} 
p^\nu p^{\nu'}$, where this term can originate from 
the longitudinal part of a bulk-to-boundary vector or axial-vector 
propagator, or from an $A_5$ particle emitted from an axial-vector current at the boundary. 
Now, the Ward identities obeyed by the four-point functions calculated above are 
\begin{equation}
l_\alpha l_\beta W_4^{\mu\nu\alpha\beta}=0 \quad \textrm{and} \quad 
p_\mu p_\nu W_4^{\mu\nu\alpha\beta}=0 \quad . \nonumber 
\end{equation}
Applying the first Ward identity to the general form for the correlator, we obtain that 
\begin{equation}
e_2 T^\mu T^\nu + g_1 p^\mu p^\nu + f_4 p^\mu T^\nu + f_3 T^\mu p^\nu= 0. \nonumber 
\end{equation} 
But this can only mean that each term in this equation separately vanishes, because 
these terms are linearly independent. Arguing similarly, one can apply the second Ward 
identity to find that the coefficients which must be zero are $e_1,e_2,f_1,f_2,f_3,f_4,
\textrm{and }g_1$. Now, what we are really after is the Lorentz singlet quantity given by 
the contraction of $W_4^{\mu\nu\alpha\beta}$ with $\eta_{\mu\nu} \eta_{\alpha\beta}$ 
(the equivalent of $W_{LLRR}(Q^2)$ and $W_{LRLR}(Q^2)$). This contraction trivially  
removes all the remaining non-zero terms, apart from the $c_1$ term, which is precisely 
the one composed entirely of transverse external propagators. Thus, we have shown that 
the only term that contributes to the scalar functions $W_{LLRR}(Q^2)$ and $W_{LRLR}(Q^2)$ 
is the one obtained by using the transverse propagators only on the external lines of 
the Witten diagrams. 

\newpage 
 

\end{document}